\documentclass{article}

\usepackage{arxiv}

\usepackage[utf8]{inputenc} 
\usepackage[T1]{fontenc}    
\usepackage{hyperref}       
\usepackage{url}            
\usepackage{booktabs}       
\usepackage{amsfonts}       
\usepackage{nicefrac}       
\usepackage{microtype}      
\usepackage{lipsum}
\usepackage{multirow}
\usepackage{graphicx}
\usepackage{tikz}
\usepackage{float}
\usepackage{tabularx}
\usepackage{enumitem}
\usepackage{array}
\usepackage{varwidth}
\usetikzlibrary{shadows.blur}
\usepackage[edges]{forest}

\title{A survey on extremism analysis using Natural Language Processing}

\author{
  Javier Torregrosa \\
  Department of Computer Science\\
  Universidad Politecnica de Madrid\\
  28051 Madrid, Spain \\
  \texttt{franciscojavier.torregrosa@upm.es} \\
  \And
  Gema Bello-Orgaz \\
  Department of Computer Science\\
  Universidad Politecnica de Madrid\\
  28051 Madrid, Spain \\
  \And
  Eugenio Martinez-Camara \\
  Andalusian Research Institute in \\
  Data Science and Computational \\
  Intelligence (DaSCI) \\
  Department of Computer Science and \\
  Artificial Intelligence\\
  University of Granada\\
  18071 Granada, Spain \\
  \And
  Javier Del Ser \\
  TECNALIA\\
  Basque Research \& Technology \\
  Alliance (BRTA) \\
  and University of the Basque \\
  Country (UPV/EHU)\\
  Alliance (BRTA) \\
  48160 Derio, Spain \\
  \And
  David Camacho \\
  Department of Computer Science\\
  Universidad Politecnica de Madrid\\
  28051 Madrid, Spain \\
  }

\begin{document}
\maketitle

\begin{abstract}
Extremism has grown as a global problem for society in recent years, especially after the apparition of movements such as jihadism.  This and other extremist groups have taken advantage of different approaches, such as the use of Social Media, to spread their ideology, promote their acts and recruit followers. The extremist discourse, therefore, is reflected on the language used by these groups texts. Natural Language Processing (NLP) represents a way of detecting this type of content, and several authors make use of it to describe and discriminate the discourse held by these groups, with the final objective of detecting and preventing its spread. Following this approach, this survey aims to review the contributions of NLP to the field of extremism research, providing the reader with a comprehensive picture of the state of the art of this research area. The content includes a description and comparison of the frequently used NLP techniques, how they were applied, the insights they provided, the most frequently used NLP software tools, descriptive and classification (based on machine learning algorithms) applications, and the availability of datasets and data sources for research. Finally, research questions are approached and answered with highlights from the review, while future trends, challenges and directions derived from these highlights are suggested towards stimulating further research in this exciting research area.
\end{abstract}

\keywords{Natural Language Processing \and Radicalisation \and Extremism \and Deep Learning}

\section{Introduction}
The rise of Social Media platforms has strengthened the interest of researchers for studying human behavior on different contexts. The potential of these platforms relies on how users behavior can be traced during or even long after they have manifested it, which facilitates the work of behavioral scientists to track them under different circumstances. This is possible thanks to the chance of crawling real time data from the users, and also to the fact that most data remains stored or published during long periods of time \cite{bayerl2014social}. Data, however, must be analyzed with an adequate approach, depending on their type (\textit{e.g.} based on interactions, text or images) and their source (\textit{\textit{e.g.}} the type of social platform).

Taking into consideration that most of the content published on the Internet is textual, it is unsurprising that one of the most frequently used approaches for online pattern extraction comes from Natural Language Processing (NLP). This discipline uses a set of computational methods for making human language accessible to computers, and more specifically for giving the computers the ability to understand and generate human language  \cite{eisenstein2019introduction}. NLP techniques are used in both academia and industry for text analysis applications, such as medicine \cite{wang2018clinical,savova2019use}, mental health \cite{calvo2017natural,stewart2021applied}, economy \cite{fisher2016natural} or crime prevention \cite{schmidt2017survey}. 

One of the area that has benefited of NLP techniques on recent years is the study of extremist discourse, particularly due to the increasing use of Social Media by different extremist groups as a speaker for disseminating their ideologies. While the first relevant extremist movements of the century (\textit{e.g.} the 11-S) took advantage of emails for their communication and organization, the growth of online platforms such as blogs, forums and finally Social Media platforms (\textit{e.g.} Twitter or Facebook) has changed the way extremists communicate, recruit and disseminate their ideas \cite{dean2012dark}. The rise of groups such as Islamic State or the Alt-right, together with their use of online Social Media platforms with different objectives \cite{jawhar2016terrorists}, has represented a threat for many countries, specially considering that extremism can facilitate the justification of violent actions to achieve a movement's agenda \cite{thomas2012responding}. This threat led different countries to finance research projects and other initiatives related to the study of the traces that extremists users left online, with the aim of identifying early behaviors to stop them before embracing violent extremism. In fact, during the worst days of the jihadist threat (between 2015 and 2018), the European Union invested in several research projects where NLP was applied to track terrorism and online extremism \cite{bouzar2018stages,fernandez2018contextual,florea2019complex,torregrosa2018risktrack}. The core of most of the initiatives aimed to counter this phenomenon, detecting and classifying extremist content that could lead people to adopt these ideologies. Machine learning techniques made a great contribution to this purpose (see, for example, Scanlon \& Gerber \cite{scanlon2014automatic}).

As stated before, the use of NLP techniques has led to several contributions focused on extremism research. After the fruitful period of research from different perspectives aimed to study and analyze the extremism phenomenon, a few systematic surveys have approached the specific relationship between NLP and extremism research. These systematic reviews can be divided in two types. The first type has analyzed NLP contributions to areas conceptually related to extremism, such as hate speech \cite{fortuna2018survey} or law enforcement \cite{edwards2015systematic}. The second type gravitates on extremism, including NLP as a key part of its identification \cite{aldera2021,Gaikwad2021}. However, reviews belong to this latter type undergo two main limitations. On the on hand, their content is restricted to the specific task of detection, not covering the rest of the whole data mining process \cite{Gaikwad2021}. On the other hand, their lack of depth when studying the NLP approaches under focus \cite{aldera2021}, missing to provide a thorough description of the diverse spectrum of techniques used in both descriptive and detection processes. 

This article aims to cover the gap left by this prior work and other similar surveys by placing an emphasis on NLP contributions to extremism analysis (including both description and classification/detection tasks), with a more comprehensive and critical approach on the different types of NLP techniques used to date. To this end, a systematic review is conducted to collect and systematically analyze the literature regarding NLP contributions on the study of extremism. This review will present a whole picture about the state of the art of this research field, both from a descriptive and a comparative approach. The first one will focus on describing the features of the articles and the content they study, whereas the second will compare their outcomes to extract useful insights for researchers. To do so, five research questions (summarized in Fig. \ref{fig:index}) are formulated to orchestrate the contributions of this review:

\begin{itemize}
    \item \textbf{RQ1. What are the current topics and contributions from NLP to extremism research?}
    
    This question aims to highlight the most relevant topics analysed by the articles, such as type of extremism or which platform is used for the research approaches, among others. This will eventually help presenting a general picture of the research field.
    
    \item \textbf{RQ2. What NLP techniques are used on extremism research?}
    
    After the screening process, the NLP techniques used by each of the articles included on the review will be extracted. After that, they will be briefly described and compared, with the aim of showing their main contributions and differences.
    
    \item \textbf{RQ3. How have NLP techniques been applied in the field of extremism research?}
    
    The different applications of the NLP techniques found in the literature will be categorized and divided depending on their approach, either the description of extremist texts or their classification. The main extremist discourse features found by the articles will be highlighted, together with the machine learning algorithms used to identify extremist texts.
    
    \item \textbf{RQ4. What NLP software tools are commonly used on extremism research?}
    
    The objective of this research question is to compare the different NLP tools (open or commercial) used by the articles reviewed. 
    
    \item \textbf{RQ5. Which publicly available datasets or datasources have authors used to conduct NLP experiments on extremism research?}
    
    This research question will approach the availability of public datasets and datasources including extremist content, to facilitate researchers their experiments.
    
\end{itemize}

\begin{figure}[h]
    \centering
    \includegraphics[width=0.65\linewidth]{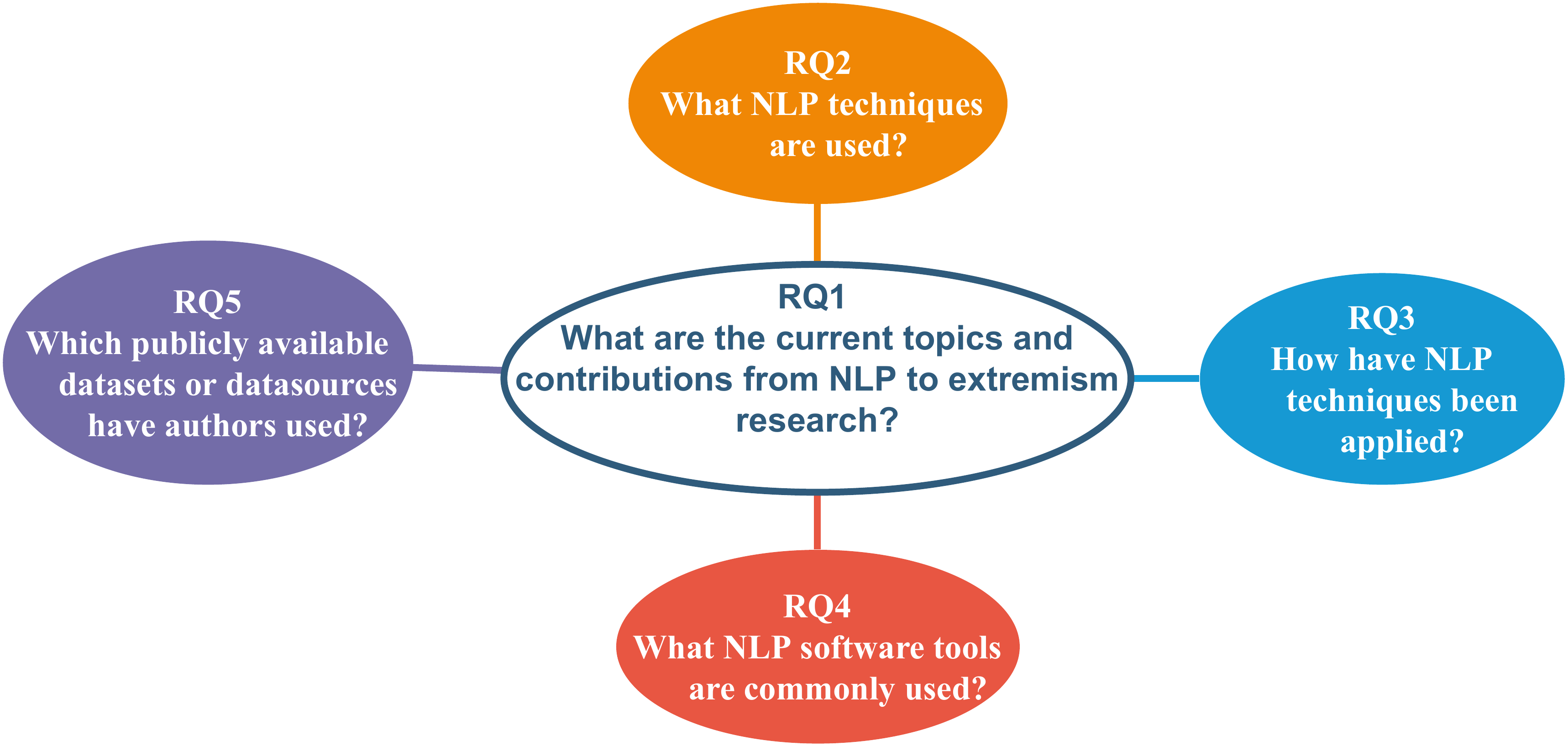}
    \caption{Summary of the research questions.}
    \label{fig:index}
\end{figure}


The main contributions of the article can be summarized in six points:

\begin{enumerate}
    \item It provides a general picture of the theoretical foundations behind the concept of "extremism", discussing its differences and similarities with other concepts that are often misused as synonyms on the literature.
    
    \item It briefly defines the concept of extremist discourse, including some key elements that are present on this type of discourses.
    
    \item It presents an updated picture of the NLP techniques (including pre-processing techniques) used on extremism research, together with an analysis and comparison of their advantages and disadvantages.
    
    \item It summarizes the different applications that NLP techniques can have on extremism research, such as discourse description and classification. The main machine learning algorithms used to identify extremist content are also highlighted.
    
    \item It presents different available tools, together with open datasets and datasources regarding extremism, which may be helpful for authors interested on conducting future experiments.
    
    \item It highlights future trends, challenges and directions of this field, regarding the conclusions extracted from the analysis. 
    
\end{enumerate}

A summary of the structure of the paper can be seen in Figure \ref{fig:index}, which is presented as follows:
Section \ref{state_of_the_art} defines what is understood as extremism, the differences among extremism and other topics and what defines the concept of extremist discourse. Section \ref{methodology} explains how the review was planned and conducted, including the inclusion and exclusion criteria, and a brief summary of the process. Section \ref{general} presents a general descriptive analysis of the outcomes of the search conducted, including the trends of publication and the main keywords associated to the articles. Section \ref{Techniques} describes and compares the different NLP techniques used by the authors. Section \ref{Aplications} focus on the applications of these techniques, dividing them in two approaches: text description and text classification, including the machine learning algorithms used for this task. Section \ref{Software} describes the NLP open datasets, datasources and tools used by the authors. Finally, section \ref{Discussion} focuses on answering the research questions and on presenting future trends, challenges and directions of the area, and presents the final conclusions.

\begin{figure}[h]
    \centering
    \includegraphics[width=1\linewidth]{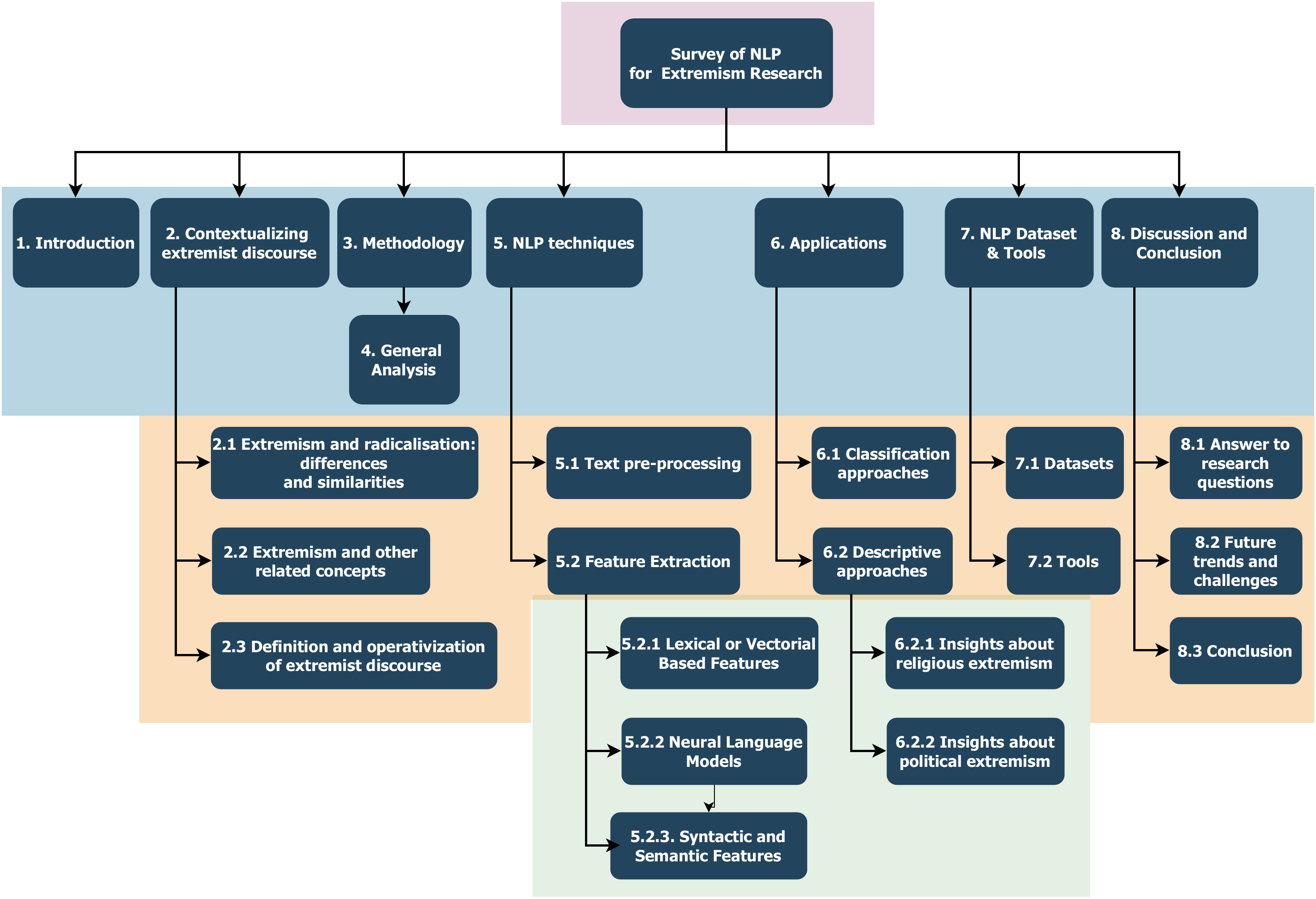}
    \caption{Structure of the overall review.}
    \label{fig:index}
\end{figure}

\section{Contextualizing the concept of extremist discourse}
\label{state_of_the_art}

The definition of extremism has traditionally led to different misconceptions in the literature, specially for authors with few background on social sciences. This section deals with the different definitions around this topic. To do so, the first subsection analysis the differences between extremism and radicalisation, two concepts that are frequently used indistinctly \cite{schmid2013radicalisation}. The second subsection briefly presents other concepts related to extremism, including definitions and relationships with it. Finally, the last subsection presents how the concept of extremism will be used in this article, including an operativization of the extremist language that will act as a framework on which the different articles reviewed can be compared.

\subsection{Extremism and radicalisation: differences and similarities}

Literature shows that extremism and radicalisation are often used as synonyms or exchangeable terms to refer the same phenomenon, which leads to the false idea that both terms mean the same. However, while authors do not usually distinguish between them from a methodological perspective, there are indeed theoretical differences that make both terms conceptually different. While there is no academic consensus about the definitions of extremism and radicalisation \cite{van2019subjectivity}, the different perspectives concerning their relationship can be summarized into three main approaches:
\begin{enumerate}
    \item \textbf{Both concepts are synonyms}. This could be related to the use of both terms on political discourse, which has transformed them in pejorative concepts that are used indistinctly \cite{schmid2013radicalisation}
    \item \textbf{Both concepts are different, but one of them is part of the other}. In this line, several articles use the concept of radicalisation as a term to refer to the psychological process previous to the involvement on terrorism and extremism \cite{schuurman2018reconsidering}.
    \item \textbf{Both concepts are different without a necessary relationship among them}. Regarding this approach, Botticher \cite{botticher2017towards} conducted a deep analysis of the historical roots of these concepts, when trying to define the differences underlying them. Essentially, the term radicalisation was born during the 18th century, as a way to define a movement against the establishment, but that is not inherently violent or positioned against democratic values. Meanwhile, the concept of extremism refers to an anti-democratic movement, and stands against "all those who do not embrace its dogmatic recipe for a transformation of society". Another reference to this article can be found in Schuurman and Taylor \cite{schuurman2018reconsidering} which highlight that radicalisation, understood in its historical context, does not necessarily imply a negative connotation of "change" of the socio-political order, while extremism does.
\end{enumerate}
   
Concerning the present review, it is necessary to have an open position towards the three different approaches. Extremism will be considered the core concept to study on this review, and therefore it will be used as a keyword instead of radicalisation (as all the social movements of interest for this article are, essentially, those against democratic values). However, due to the misconception or confusing use of both terms in the literature, both radicalisation and extremism will be used as keywords to conduct the search on the databases during the article gathering process. Through this decision we will be able to include articles from authors considering both terms as synonyms and those using one as part of the other. 

\subsection{Extremism and other related concepts}

Similarly to the terms extremism and radicalisation, there are other concepts that are currently confusing on their use in the context of extremism research. While some of these terms are quite related, they do not share the same theoretical definition. 

Table \ref{tab:concepts_related_extremism} includes some of these terms, their definition, their difference with the concept of extremism and an example from the literature regarding them. Taking into account that the main characteristic to classify a movement as extremism is that it goes against democratic values, we can find three different types of concepts related to extremism in this table. The first two terms (supremacism and sectarianism) are actually subtypes of extremism since they are both different types of ideological movements that try to suppress or limit certain fundamental democratic values of other social groups. When these ideological movements against democratic values use violence to try to achieve their objectives, it could be said that they constitute a type of Terrorism (third term in the Table). Finally, the last three terms shown in the Table (polarization, fundamentalism and nationalism), although are related to extremism, do not necessarily share its main characteristic of going against democratic values.  

There are other concepts that, appearing related to extremism, are just manifestations of the violence and discrimination underlying this concept. Some examples could be hate speech \cite{olteanu2018effect}, racism \cite{fuchs2016racism} or stalking/cyber-stalking \cite{kruglanski2020psychology}. The creation of fake news \cite{spohr2017fake} and its relationship with extremism currently represents another rising problem that has attracted the attention from researchers. 

\begin{table}
\centering
\caption{Concepts, definitions and distinction from extremism.}
\begin{tabular}{|l|p{4cm}|p{4cm}|p{3.2cm}|}
\hline
\textbf{Concept} &
  \textbf{Definition} &
  \textbf{Distinction from extremism} &
  \textbf{Example of the concept}\\ \hline

\textit{Supremacism} &
  Ideology that assumes that one group is naturally superior to another one, due to their race, sex, economic status, nation, etc. \cite{schaefer1990racial} &
  Could be a subtype of extremism, as supremacist groups are contrary to the existence of equal rights.
  &
  White supremacist movement \cite{kantrowitz2015ben} \\ \hline
 
 \textit{Sectarianism} &
  Form of discrimination between groups based on a specific factor. For years, it was limited to religion, but nowadays this concept is technically similar to supremacism. \cite{phillips2015sectarianism} &
  Would be a subtype of extremism as Supremacism, and it is contrary to the existence of equal rights.
  &
  Conflicts between Nationalists and Unionists in Northern Ireland \cite{cairns1998conflict} or political disparity between Shia and Sunni Muslims \cite{wehrey2017beyond}
  \\ \hline
  
\textit{Terrorism} &
  Systematic use of violence, propaganda and fear towards and specific population to achieve ideological objectives. \cite{lopez2016boko} &
  Always implies violence, while extremism does not necessarely use it. However, both are against one or more fundamental values of a society.
  &
  IRA in Ireland/North Ireland \cite{pruitt2007readiness}, ETA in Spain \cite{shepard2002eta}, FARC in Colombia \cite{saab2009criminality}, The Islamic State \cite{roy2017jihad} or Al'Qaeda \cite{burke2004qaeda} \\ \hline
  
  \textit{Polarization} &
  Ideological movement towards a more extreme point of view in whatever direction is indicated by the member's predeliberation tendency \cite{sunstein1999law}
  &
  It is not necessarely violent or against fundamental values of a society, as occurs with radicalization.  
  &
  Political or ``partisan" polarization \cite{prior2013media} \\ \hline

\textit{Nationalism} &
  Ideology based on the nodal point ``nation", on which a community is tied to a certain space, and that is structured through the opposition between the nation and different outgroups. \cite{de2017populism} &
  Does not necessarily imply a negative connotation. When it turns extremist, it would convert to supremacism.
  &
  Catalonia, Scotland and Canada have some renowned political movements related to nationalism \cite{keating1996nations} \\ \hline

\textit{Fundamentalism} &
  Tendency to follow literally certain dogmas or ideologies from the "fundamental" and unchangeable practices of the past. As sectarianism, it has a religious connotation. \cite{hunsberger1995religion} &
  Is not necessarely violent or against democratic values. 
  &
  The ``Amish" (example of christian fundamentalist group) \cite{hill2005psychology}. \\ \hline

\end{tabular}
\label{tab:concepts_related_extremism}
\end{table}

\subsection{Definition and operativization of extremist discourse}
\label{operativization}

Until now, it has been presented a distinction between the concepts of radicalization and extremism, choosing extremism as a key concept to justify the aims of this article. Also, extremism has been compared with other concepts that tend to appear. As has been stated, this term can have different meanings depending on the approach considered by the author, and this is why its relevant to establish a clear definition to work with. In this review, our definition of extremism will be "\textit{an ideological movement, contrary to the democratic and ethical values of a society, that uses different methods, including violence (physical or verbal) to achieve its objectives}". 

Following this definition, a second step would be to clarify what does this article means when it refers to extremist discourse. While it could be addressed as "\textit{the use of language held by people when expressing their extremist views}", it shall be notice how authors have highlighted several features that characterizes an extremist narrative from a regular discourse. These features, derived from different authors \cite{ashour2010online,bennett2011war,fortuna2018survey,sakki2016discursive,torregrosa2020linguistic}, can be summarized as follows:
\begin{itemize}
    \item \textbf{Types of extremist narrative:} there are several ways on which extremist narratives try to justify their vision and objectives. Ashour \cite{ashour2010online} divided these narratives into five categories: political, historical, socio-psychological, instrumental and theological/moral. 
\begin{itemize}    
\item \textbf{Political:} the discourse includes references to grievances from one or more group towards other group. 
\item \textbf{Historical:} legitimization of the political grievance narratives through the use of historical examples and similes. 
\item \textbf{Socio-psychological:} glorification of acts against the system, both violent or not.
\item \textbf{Instrumental:} justification of the violence and "self-defense" as a way of reaching objectives.
\item \textbf{Theological/moral:} legitimization of actions or reactions against political grievance or social oppression through religion, morality or ethics.
\end{itemize}
    \item \textbf{Linguistic style:} the narrative styles or topics previously mentioned are built based on a specific vocabulary and style, that helps extremists structuring their discourse. Several articles have found differences on the linguistic style from radical and extremist texts compared to a regular sample of texts \cite{cohen2014detecting}. For example, the higher use of first and third person plural pronouns, a more negative tone or the use of more words related to negative topics are common in these texts  \cite{torregrosa2020linguistic}.  
    \item \textbf{Use of discursive resources such as hate speech, otherness or war narrative:} extremist texts tend to use discursive resources to justify their actions and ideas towards others. Some of these techniques have been deeply studied, such as hate speech \cite{fortuna2018survey}, otherness \cite{sakki2016discursive} or the use of war terminology to create "enemies" and a "call to action" on others \cite{bennett2011war}. 
\end{itemize}

On this point, both the definition and operativization of extremist discourse have been stated. This type of discourse is characterized by the use of specific narratives, an aggressive and polarized linguistic style and several techniques oriented to justify a feeling of superiority or inferiority towards another group. The next sections of this article will review how authors have used NLP to detect and describe the extremist discourse on Social Media, and the outcomes they have reached.

\section{Methodology}
\label{methodology}

This section describes the process carried out to conduct the survey of the articles that apply NLP to extremism research. This process was conducted through a systematic approach, extracting all the articles from four databases: Scopus, ScienceDirect, IEEE Xplore and Web of Science.

Concerning the thesaurus used for the search, it was decided to use both the terms extremism and radicalisation on the search. The reason behind this decision was that, as stated before, it is quite common that authors miss-use these concepts as synonyms \cite{botticher2017towards,schmid2013radicalisation}. 

Second, while the thesaurus "Natural Language Processing" was included, it was also decided to expand the search with different subtopics, such as "Sentiment analysis", "Topic detection" and "Semantic analysis". Eventually, and due to the recent contributions from the field of deep learning to natural language processing \cite{young2018recent}, it was decided to include also the subtopic "Deep learning" to the search.

Therefore, the thesaurus finally included on the searching process are presented below:
\\

\textit{("Natural Language Processing" OR "Sentiment Analysis" OR "Topic Detection" OR "Semantic Analysis" OR "Deep Learning") AND ("Extremism" OR "Radicalization")}
\\

No time limits were selected when conducting the review, meaning that the articles could be published any year. The extraction was conducted in January 2021. 729 documents were found on the different databases. Table \ref{screen_databases} shows the distribution of articles found per database. After deleting the duplicates and the non-scientific articles (e.g. indexes), 675 articles remained on the survey.

\begin{table}
\centering
\caption{Articles extracted from the different databases that apply NLP to extremism research.}
\begin{tabular}{|l|r|}
\hline
\textbf{Datasource} & \textbf{No. Articles} \\ \hline
ScienceDirect & 95 \\ \hline
Scopus & 573 \\ \hline
Web of Science & 41 \\ \hline
IEEE Xplore & 20 \\ \hline
\end{tabular}
\label{screen_databases}
\end{table}

After the searching process, a general screening of the articles was conducted. This screening process included checking the title, the abstract and the methodology to find out if the articles accomplished the inclusion criteria of the review. This criteria can be summarized as:

\begin{enumerate}
    \item The documents shall empirically apply NLP to extremism description or classification. 
    \item The analysis conducted on the documents shall be quantitative.
    \item The documents shall clearly state the NLP techniques they use to conduct the analysis. 
    \item The documents shall present a clear methodology, including all the scores and the process they followed to conduct the analysis.
    \item The article shall be written in English
\end{enumerate}

After this general screening, 70 documents remained on the review. Next, a more exhaustive review was conducted over those 70 articles, reading the content of the document and excluding the ones not accomplishing the criteria presented above. After the second screening, 6 articles were discarded. The rest of the articles (a total of 64) were finally included for the review. 


\section{General descriptive analysis of the articles}
\label{general}

This section presents a general descriptive analysis of the articles finally included on the review. Firstly, a general introduction is presented where the publishing years and the types of extremism detected are reviewed. Then, to identify the most relevant topics related to NLP that deal with the selected articles, a textual analysis has been performed using their indexing keywords. This description will also be used to structure the following sections of the paper, as it shows a general picture about the main topics addressed by the reviewed papers.

\begin{figure}[h]
    \centering
    \includegraphics[width=1\linewidth]{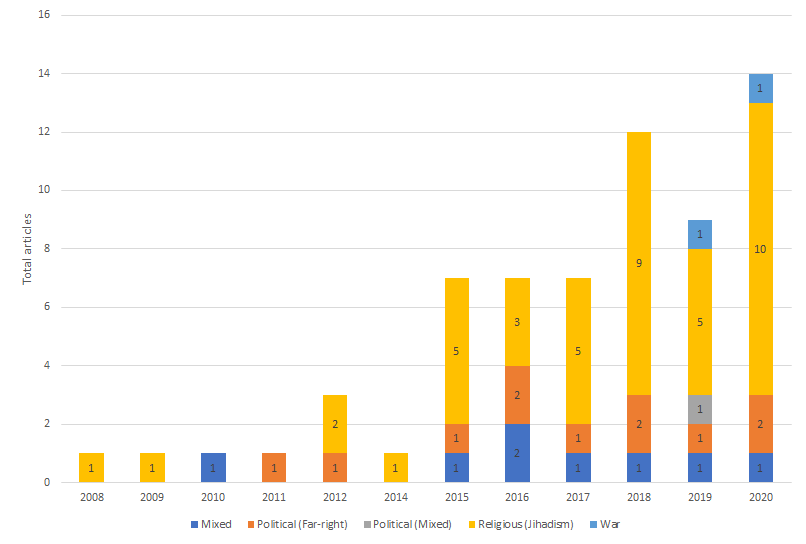}
    \caption{Type of extremism addressed by the articles included in the survey}
    \label{fig:type_radical}
\end{figure}

Analyzing the timeline of the publications reviewed and the type of extremism addressed, it can be seen that the interest for conducting research works applying NLP to study extremism has been increasing during recent years, as shown in Fig \ref{fig:type_radical}. This in turn supports the ideas presented on the introduction of this article: most of the articles were published during or after 2015, which overlaps the time lapse when ISIS was more active.

Besides, as Fig \ref{fig:type_radical} confirms, the most frequently addressed type of extremism in the articles is jihadi extremism, with a significant gap with the rest of types. In general terms, there are 5 types of extremism approached in the literature reviewed: religious (all of them concerning jihadism), political (far-right) political mixed (concerning far right/far left), war (concerning conflicts in different countries, such as Afghanistan), and mixed (studying both religious and political extremism). Since 2015, the studies approaching NLP and extremism grew year by year has, having a substantial increase. In this last period, while jihadi extremism has attracted more interest, political extremism remains relatively steady. Therefore, it could be concluded that the two predominant types of extremism analyzed religious and political.

\begin{figure}[h]
    \centering
    \includegraphics[width=0.65\linewidth]{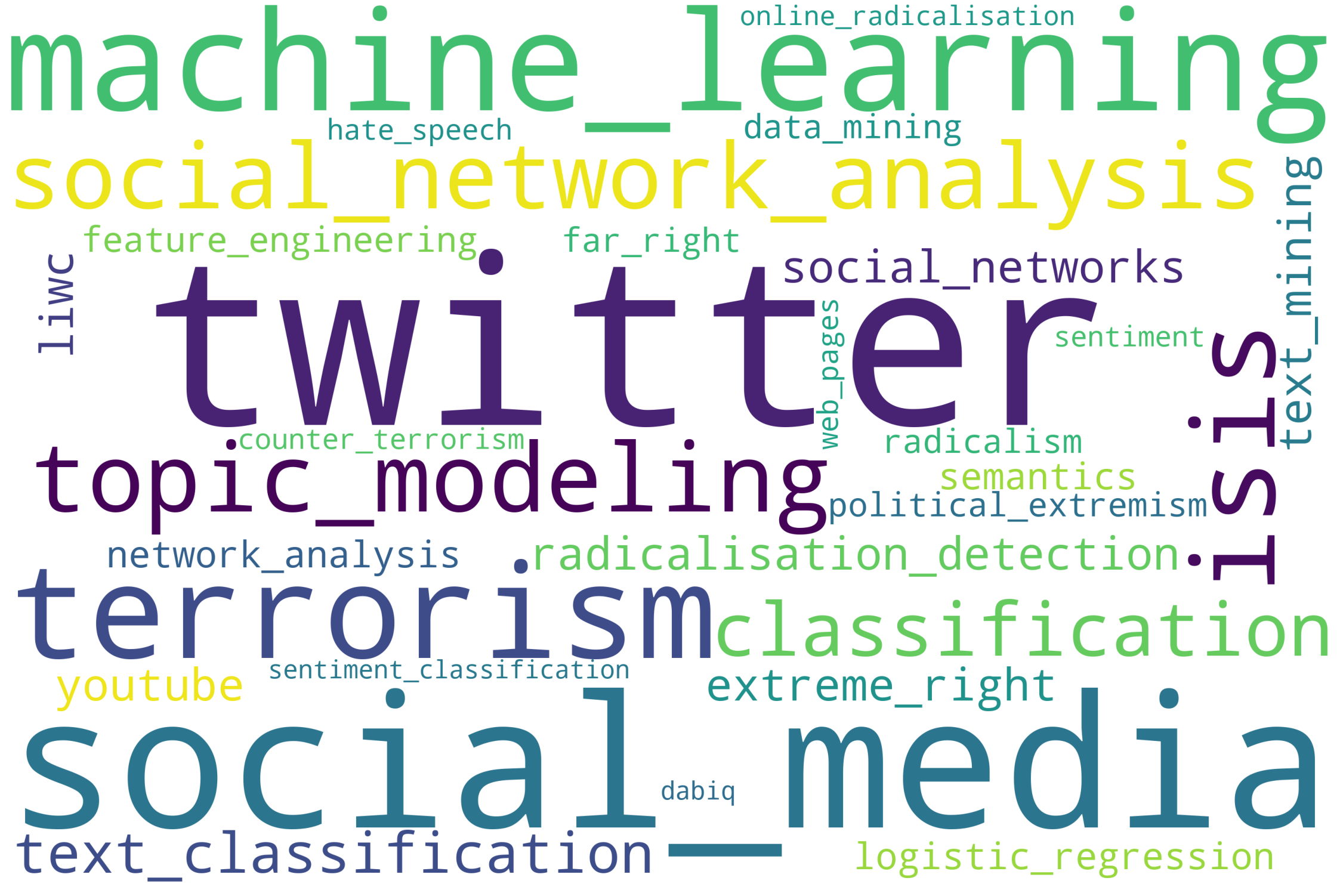}
    \caption{Word cloud of keywords extracted from the articles analysed}
    \label{fig:WordCloudKeywords}
\end{figure}

Continuing the preliminary analysis to determine the more common topics associated with the thesaurus used on the search on the articles, a textual analysis of the keywords related to the reviewed articles has been performed. For this purpose, Fig. \ref{fig:WordCloudKeywords} shows a word cloud with the top 30 of the most frequently used keywords by the articles (keywords used as thesaurus were excluded from the count). 

As can be seen, there are different keywords than can be grouped under 4 similar concepts. First, different NLP techniques are mentioned on the keywords (e.g. topic modeling, sentiment classification or semantics). Second, the source of the data analyzed also appears frequently on the articles keywords (e.g. Twitter, social media, YouTube, web pages, or Dabiq, a jihadi magazine), as well as specific tools that can be used (e.g. LIWC). Third, different keywords related to extremism are mentioned (e.g. terrorism, ISIS, far-right, extreme right, hate speech, online radicalisation, or radicalism). Finally, some keywords are related to the methods applied, including classification techniques, to detect extremist content (machine learning, classification, data/text, logistic regression or feature engineering). It should be mentioned that, while not the objective of this survey, Social Network Analysis (SNA) appeared as one of the most used keywords (both Social Network Analysis, Social Networks and Network Analysis), and one concurrent approach when conducting NLP analysis.

\section{NLP techniques for extremisn research}
\label{Techniques}
The main objective of NLP techniques is to transform free text into structured data by capturing its lexical, syntactic and semantic information to acquire or infer new knowledge. Considering this, the NLP process can be divided into two main phases: text pre-processing (simplifying and preparing the text for its analysis) and feature generation (transforming the text into a structured data representation suitable to be used by the different algorithms or methods of analysis). According to this division, the following subsections present a detailed analysis of the techniques used by the articles reviewed for each of those specific phases.

\subsection{Text pre-processing}
\label{pre-processing Techniques}
The pre-processing of textual data is a key part of NLP, as it helps to identify and establish the fundamental units that will be assessed during the analysis \cite{kannan2014preprocessing,vijayarani2015preprocessing}. This process includes a set of techniques that allow NLP algorithms to compute and analyse words, simplifying and preparing the text for its analysis. The pre-processing techniques mentioned on the articles are:

\begin{itemize}
    \item \textbf{Tokenization:} Process of dividing a sentence in smaller units (tokens), such as words.
    \item \textbf{Cleaning:} Removal of strange or non-informative characters of the text URLs, such as symbols, punctuation, hashtags or other special characters.
    \item \textbf{Stop-words}: Removal of words that occurs frequently and do not carry relevant information in most contexts (such as articles or prepositions). 
    \item \textbf{Lowercasing:} Process of lowercasing the capital letters (some NLP algorithms do not  discriminate between lower and uppercase).
    \item \textbf{Lemmatization:} Process of reducing inflected words to their roots (lemma).
    \item \textbf{Stemming:} Process of erasing prefixes or suffixes from a word to obtain its stem.
\end{itemize}

Table \ref{tab:pre-processing} shows all the articles reviewed that explicitly mention using one or more of these pre-processing techniques. Tokenization is an essential task in natural language processing used to break up a string of words into semantically useful units called tokens, so all the articles, even if not explicitly mentioned, perform this pre-processing as part of it NLP process. 

The rest of the preprocessing tasks may or may not be performed depending on the application to be carried out. As shown in Table \ref{tab:pre-processing}, most of the works reviewed apply cleaning processes to the texts to eliminate unnecessary characters and words, while few of them use the process of lowercasing. However, in the case of the process of transformation of the words to their stems or lemmas, it can be concluded that stemming is the most used technique. The aim of both techniques is to reduce the words to a common base form. As opposed to stemming, lemmatization does not simply chop off inflections. Instead it uses lexical knowledge bases to get the correct base forms of words, and it is more expensive computationally, which may be the reason for its minor application.

\begin{table}[h]
\centering
\caption{NLP pre-processing techniques applied in reviewed articles.}
\begin{tabular}{|l|r|p{8cm}|}
\hline
\textbf{Pre-processing techniques} & \textbf{Percentage Use} & \textbf{Articles explicitly mentioning them} \\ \hline
 Tokenization & 100\% & All the articles use tokenization. \\ \hline

Stop-words & 35.93\% &
  \cite{masood15using,abddetecting2020,nouh2019understanding,johnston2017identifying,ahmad2019detection,sharif2019empirical,hartung2017identifying,mariconti2019you,o2012analysis,ben2016hate,saif2017semantic,mirani2016sentiment,rehman2021understanding,rekik2020recursive,sharif2020detecting,heidarysafa2020women,rekik2019violent,zahra2018framework,fernandez2018understanding,kinney2018theming,sabbah2015hybridized,alghamdi2012topic,bermingham2009combining} \\ \hline
  
Filtering &  37.5\% &
  \cite{abddetecting2020,nouh2019understanding,ahmad2019detection,hartung2017identifying,o2012analysis,saif2017semantic,mirani2016sentiment,rehman2021understanding,rekik2020recursive,sharif2020detecting,heidarysafa2020women,rekik2019violent,zahra2018framework,fernandez2018understanding,kinney2018theming,sabbah2015hybridized,ottoni2018analyzing,araque2020approach,gomes2017profiling,agarwal2015using,torregrosa2020analyzing,alizadeh2019psychology,bisgin2019analyzing,hall2020machines} \\ \hline
  Lowercasing &  9.37\% &
  \cite{zahra2018framework,fernandez2018understanding,nouh2019understanding,araque2020approach,hall2020machines,kursuncu2019modeling} \\ \hline
Lemmatization &  4.68\% &
  \cite{ottoni2018analyzing,nouh2019understanding,figea2016measuring} \\ \hline
Stemming &  20.31\% &
  \cite{bermingham2009combining,alghamdi2012topic,o2012analysis,sabbah2015hybridized,mirani2016sentiment,zahra2018framework,sharif2019empirical,mariconti2019you,masood15using,rehman2021understanding,figea2016measuring,stankov2010contemporary,yang2011social} \\ \hline

\end{tabular}%
\label{tab:pre-processing}
\end{table}

\subsection{Feature Extraction}
\label{FeatureGeneration}

After the pre-processing of the textual data, different text mining techniques are used to transform tokens into structured data by capturing its lexical, syntactic and semantic information. These structured data can be eventually used as input for the different algorithms to acquire or infer new knowledge. Table \ref{NLP_technique_Summary} presents all the techniques mentioned on the review, together with the articles included on the review that have apply them as part of their methodological approach. These techniques can be grouped into 3 different categories according to the type of linguistic information captured, which are explained below in detail in the following sections. A first descriptive analysis of the techniques is conducted for each of these subsections. Afterwards, a comparative analysis of these techniques is carried out within the area of research on extremism, highlighting the advantages and disadvantages of each technique within this specific domain.

\begin{table} [h]
\centering
\caption{Summary of NLP techniques for feature generation using in the articles reviewed.}
\begin{tabular}{|p{3.1cm}|p{2.5cm}|r|p{6cm}|}
\hline
\textbf{Approach} &
  \textbf{NLP technique} &
  \textbf{\begin{tabular}[c]{@{}l@{}} Percentage Use \end{tabular}} &
  \textbf{Articles} \\ \hline
\multirow{6}{*}{Lexical or Vectorial} &
  N-grams &
  28,12\% &
  \cite{de2020radical,rehman2021understanding,sharif2019empirical,kinney2018theming,masood15using,kim2017empirical,hartung2017identifying,saif2017semantic,ben2016hate,prentice2012language,rekik2019violent,rekik2020recursive,fernandez2018understanding,sharif2020detecting,abddetecting2020,kursuncu2019modeling,nouh2019understanding,hall2020machines} \\ \cline{2-4} 
 &
  Dictionaries &
  37.5\% &
  \cite{scrivens2020measuring,alizadeh2019psychology,devyatkin2017exploring,mirani2016sentiment,saif2016role,bisgin2019analyzing,rowe2016mining,scanlon2015forecasting,gomes2017profiling,johnston2017identifying,johnston2020identifying,ottoni2018analyzing,hall2020machines,saif2017semantic,abdelzaher2019systematic,dillon2020comparison,klein2019online,owoeye2018classification,rekik2019violent,rekik2020recursive,torregrosa2020analyzing,wei2018detecting,fernandez2018understanding,smith2020detecting} \\ \cline{2-4} 
 & TF &
  50\% &
  \cite{abdelzaher2019systematic,agarwal2015using,ben2016hate,bisgin2019analyzing,chen2008sentiment,de2020radical,dillon2020comparison,figea2016measuring,hartung2017identifying,kinney2018theming,klein2019online,macnair2018changes,owoeye2018classification,owoeye2019classification,rekik2019violent,rekik2020recursive,rowe2016mining,scanlon2015forecasting,scrivens2015sentiment,scrivens2020measuring,torregrosa2020analyzing,wei2016identification,wei2018detecting,alizadeh2019psychology,fernandez2018understanding,smith2020detecting,bermingham2009combining,araque2020approach,kursuncu2019modeling,prentice2012language,alghamdi2012topic,devyatkin2017exploring,stankov2010contemporary} \\ \cline{2-4} 
 &
  TF-IDF &
  23.43\% &
  \cite{alghamdi2012topic,ahmad2019detection,heidarysafa2020women,mariconti2019you,o2012analysis,rehman2021understanding,sabbah2015hybridized,sharif2019empirical,sharif2020detecting,yang2011social,zahra2018framework,abddetecting2020,kim2017empirical,masood15using,nouh2019understanding} \\ \cline{2-4} 
 &
  Dichotomous appearance &
  1.56\% &
  \cite{wadhwa2015approach} \\ \cline{2-4} 
 &
  Log-likelihood &
  3.12\% &
  \cite{stankov2010contemporary,prentice2012language} \\ \hline

\multirow{3}{*}{Neural Lenguage Models} &
  Word2Vec &
  9.37\% &
  \cite{abddetecting2020,araque2020approach,johnston2020identifying,kim2017empirical,kursuncu2019modeling,masood15using,nouh2019understanding,ottoni2018analyzing} \\ \cline{2-4} 
 &
  FastText &
  4.68\% &
  \cite{ahmad2019detection,araque2020approach,devyatkin2017exploring} \\ \cline{2-4} 
 &
  GloVe &
  3.12\% &
  \cite{araque2020approach,gomes2017profiling} \\ \hline

\multirow{8}{*}{Sintantic and Semantic} &
  Part-of-speech &
  25\% &
  \cite{devyatkin2017exploring,owoeye2018classification,mariconti2019you,masood15using,wignell2018natural,macnair2018changes,figea2016measuring,skillicorn2015empirical,scrivens2016sentiment,scrivens2018searching,weir2016positing,de2020radical,owoeye2019classification,scrivens2015sentiment,sikos2014authorship,yang2011social} \\ \cline{2-4} 
   &
  NER &
  7.81\% &
  \cite{bisgin2019analyzing,saif2017semantic,saif2016role,fernandez2018contextual,hartung2017identifying} \\ \cline{2-4} 
 &
  LSF &
  4.68\% &
  \cite{kim2017empirical,masood15using,hartung2017identifying} \\ \cline{2-4} 
 &
  Parse trees &
  1.56\% &
  \cite{sikos2014authorship} \\ \cline{2-4} 
  &
  LDA &
  15.62\% &
  \cite{bisgin2019analyzing,scanlon2015forecasting,ottoni2018analyzing,hall2020machines,saif2017semantic,kursuncu2019modeling,heidarysafa2020women,alizadeh2019psychology,kinney2018theming,kim2017empirical} \\ \cline{2-4} 
 &
  NMF &
  4.68\% &
  \cite{heidarysafa2020women,o2015down,o2012analysis} \\ \cline{2-4} 
 &
  Sentiment scoring &
  37.49\% &
  \cite{wignell2018natural,chen2008sentiment,saif2017semantic,hartung2017identifying,masood15using,heidarysafa2020women,hall2020machines,owoeye2018classification,macnair2018changes,figea2016measuring,scrivens2016sentiment,scrivens2018searching,weir2016positing,owoeye2019classification,scrivens2015sentiment,mirani2016sentiment,rowe2016mining,dillon2020comparison,torregrosa2020analyzing,araque2020approach,scrivens2020measuring,wei2016identification,bermingham2009combining,ahmad2019detection} \\ \cline{2-4} 
 &
  Semantic tagging &
  12.50\% &
  \cite{wignell2018natural,saif2017semantic,saif2016role,fernandez2018contextual,ottoni2018analyzing,devyatkin2017exploring,abdelzaher2019systematic,prentice2012language} \\ \cline{2-4} 

&
  Word/sentence length &
  7.81\% &
  \cite{stankov2010contemporary,yang2011social,sikos2014authorship,weir2016positing,scrivens2018searching} \\ \cline{2-4} 
 &
  Use of emoticons &
  3.12\% &
  \cite{agarwal2015using,wei2016identification} \\ \cline{2-4} 
 & 
  Use of punctuation &
  3.12\% &
  \cite{sikos2014authorship,yang2011social} \\ \hline 
\end{tabular}%
\label{tab:NLP_technique_Summary}
\end{table}

\subsubsection{Lexical or Vectorial Based Features}
\label{LexicalVectorial}

The tokens extracted from the pre-processing phase have to be transformed into more complex data structures representing a final textual features to be further processed. For this purpose, different techniques of text representation modeling can be applied. Vector Space Models (VSM) \cite{turney2010frequency} is one of the most widely text representation used in classical NLP approaches. The idea of the VSM is to represent each text or document in a collection as a point in a space (a vector in a vector space) based on the token extracted. After the tokenization process, the first step to generate this type of representation consists on defining the weighting technique to compute the tokens (terms) appearance's frequency in a text. The articles reviewed mention several different techniques to generate this vector representation:

\begin{itemize}
\item \textbf{N-grams:} tokens of size 1 are obtained from pre-process the free texts, which means that represents only one word. However, sentences generally contain compound terms (such as living room or coffee machine) formed by several words with a single meaning. The use of grouping multiple tokens together to represent that inherent meaning can be very beneficial for NLP subsequent tasks, and this is what n-grams models provide \cite{sidorov2012syntactic}. A uni-gram is any single element of the text, while a bi-gram or a tri-gram is composed by two or three elements, respectively, that appear sequentially on the text. Skip-gram is a special version of n-gram, as it works the same way, but considering tokens that are not necessarily juxtaposed on the text. Therefore, an analysis based on n-grams consider n elements as a single token. One of the main advantages of this approach is that high ``n" sizes help providing  context for words \cite{fortuna2018survey}. Table \ref{tab:ngram_type} summarizes which type of n-gram used the articles reviewed, where the unigrams are not shown since, as mentioned above, they would be 1-sized tokens that have been already obtained through the pre-process techniques. 

\begin{table}
\centering
\caption{Type of n-gram used on the article reviewed. }
\resizebox{\textwidth}{!}{%
\begin{tabular}{|l|l|p{8cm}|}
\hline
\textbf{N-gram type} & \textbf{Percentage Use}  &
  \textbf{Articles using it} \\ \hline
Bi-gram & 15.62\% &
  \cite{de2020radical,rehman2021understanding,sharif2019empirical,kinney2018theming,masood15using,kim2017empirical,hartung2017identifying,saif2017semantic,ben2016hate,prentice2012language} \\ \hline
Bi-gram + Tri-gram & 6.25\% &
  \cite{rekik2019violent,rekik2020recursive,fernandez2018understanding,sharif2020detecting} \\ 
  \hline
Bi-gram + Tri-gram + Skip-gram & 4.68\% &
\cite{abddetecting2020,kursuncu2019modeling,nouh2019understanding} \\ 
  \hline
Tri-gram + Tetra-gram + Penta-gram & 1.56\% &
  \cite{hall2020machines} \\ \hline
\end{tabular}%
}
\label{tab:ngram_type}
\end{table}

    \item \textbf{Dictionaries:} uses pre-established lists of lexicons (words or sentences) to filter or group the pre-processed tokens. Therefore, any term found inside the lexicon is considered as a final token to generate the final text representation. Dictionaries can also group the frequency of terms as a whole token, thus calculating the frequency of occurrence of a dictionary itself. The main advantage of the dictionaries is that they capture concepts defined by different terms. However, they are also very vulnerable to words not previously included on the lexicon.
    
    \item \textbf{Term frequency (TF):} is the more basic weighting technique in NLP, and consists on the raw sum of the apparition of each token found in the text. It can be represented as \textit{tf(t, d)}, on which \textit{t} is the number of times a token appears on the document \textit{d}.
    
    \item \textbf{Term Frequency - Inverse Document Frequency (TF-IDF):} is an evolution of the aforementioned TF. While the TF just sums the frequency of occurrence of a token in a text, TF-IDF also divides it by the frequency of occurrence of a word in the whole corpus. When a word is more frequent in a text than in the whole corpus, it means that this word is relevant for the text, and therefore it has a higher score \cite{chen2008using}. It is useful to discriminate between relevant words and words with no relevant meaning, such as stop-words \cite{fortuna2018survey}.

    \item \textbf{Dichotomous appearance:} this technique does not consider the frequency of a token, but instead the presence/not presence of a token. Therefore, it is computed as 0 if the term does not appear, and 1 if the term appears.

    \item \textbf{Log-likelihood:} this association metric \cite{dunning1993accurate} is used to compute the significance of co-occurrence of two variables (for example, two tokens, a token with the group used to classify, etc.). Therefore, this technique is not focused on the frequency of a single token, but in the frequency of two conditions appearing together, which may include one or two tokens.
    
\end{itemize}

Focusing on Table \ref{NLP_technique_Summary}, the first point to note is the high use of N-grams and dictionary techniques, exceeding 25\% in both cases. This is due to the fact that, from the text pre-processing phase, tokens of size 1 are obtained representing the text, and in many cases, before applying more complex techniques that transform such tokens into complex data structures, it is beneficial to apply some basic NLP techniques. These techniques allow to group or filter the tokens by aggregating them a first level of lexical information. 

The major advantage provided by the N-grams approach is that it is independent from the text. This means that all the text can be vectorized using these techniques, no matter if they appear on a lexicon or not. This is specially useful when applying NLP to extremism research, as texts usually combine terms in different languages. However, this versatility also has a handicap: the terms vectored may have no relevant meaning for the researcher, and therefore extra work shall be conducted here to identify which terms are relevant or not.

On the other hand, the use of dictionaries is helpful to detect and classify tokens into meaningful psycho-linguistic categories \cite{fernandez2018understanding,figea2016measuring}. This is a great advantage in the field of extremism research, taking into consideration  the psychological background that motivates extremist behaviour. In fact, one of the main dictionaries' based tool, LIWC, was born with the aim of conducting psychological research from texts, and has been frequently applied to extract psychological insights and extremist slang from extremist texts \cite{torregrosa2020linguistic}. However, they require a previous effort from the researchers to prepare the lexicons or to adapt them to other languages \cite{sikos2014authorship}. This last point is specially relevant in the case of jihadi extremism, as texts usually combine Islamic terminology (written in Arabic) with different languages \cite{sikos2014authorship}.

Continuing with the analysis of the Vectoral Space Models applied in the reviewed articles, TF and TF-IDF are the most used techniques. The second is a evolution from the first one, using IDF to eliminate the common terms from the language that are not relevant to categorize texts (in this case, extremist content). Taking into consideration that several articles from the review conduct filtering pre-processing techniques to eliminate irrelevant terms (such as stop-words), there is not a huge difference among them concerning the extremism research field. The main advantage of these techniques is their simplicity and broad use, which lead them to be the most commonly applied  techniques. But they have the great disadvantage that they do not provide semantic information about the terms.

Dichotomous appearance was only used in one article. While it presents a clear advantage (it is quite easy to implement), it has one main disadvantage: as stated on the previous section, some terms are used with different semantic meanings in regular and extremist texts \cite{fernandez2018contextual,gomes2017profiling,saif2016role,wei2018detecting}. Analyzing only the apparition of a term can be poorly informative for the researcher. Finally, Log-likelihood can be used for analysing association among terms, which allows to provide more contextual information, but it is still a very little used technique within the extremist field of study.

A brief summary of the advantages and disadvantages of all these techniques appears on table \ref{tab:token_featurization_comparison}:

\begin{table}[h]
\centering
\caption{Comparison of Vector Space Models based techniques to generate features in the articles reviewed.}
\begin{tabular}{|c|p{5.3cm}|p{5.3cm}|}
\hline
\textbf{Technique} &
  \multicolumn{1}{c|}{\textbf{Advantages}} &
  \multicolumn{1}{c|}{\textbf{Disadvantages}} \\ \hline
N-grams &
  -Able to keep semantic information. &
  -Captures basic semantic information. \\  
 &
  -High versatility, due to its independence from the text (useful for multi-language texts) &
  -The tokens detected may not have interest for the researcher. \\ \hline
Dictionaries &
  -Useful to conduct psycho-linguistic meaningful analysis &
  -Low versatility (vulnerable to changes on the language and word structure). \\  
\multicolumn{1}{|l|}{} &
  -Useful to detect and classify specific slang and terminology. &
  -Highly dependent on the lexicons included. \\ \hline
  TF/TF-IDF &
  -Simple and widely used.&
  -Not capture semantic context information. \\ & & -TF needs a previous stopwords filtering. \\ \hline
Dichotomous appearance &
  -The simplest technique. &
  -Not capture semantic context information. \\ \hline
Log Likelihood &
  -Captures information of association among terms. &
  -Few applied in the area. \\ \hline

\end{tabular}%
\label{tab:token_featurization_comparison}
\end{table}

\subsubsection{Neural Language Models (Word embedding)}
\label{embeddings}

Techniques based on Neural Models include a set of methods that transform tokens obtained from the pre-processing phase into meaningful vectors through the use of neural networks, allowing to capture the relationship among them \cite{levy2014dependency} and, therefore, information about words semantically related. In recent years, the application of these models in the field of extremism research have gained more relevance, as they are useful to keep information about the semantic meaning of terms. This is, precisely, the advantage of this type of models to extract textual features compared with the classics models seen in the previous section. This aspect is specially relevant when applied to classification tasks and the use of deep learning to identify extremist content \cite{johnston2020identifying,johnston2017identifying}. The most common Neural Models found on the reviewed articles are:

\begin{itemize}

    \item \textbf{Word2Vec:} allows to predict words depending on the context, maintaining the semantic meaning of the sentence. To do so, the model creates a vector related to each word through the use of a single layer neural network, which can be interpreted as a space. The words that are more likely to appear together on the text will appear closer on that space, therefore sharing semantic context \cite{mikolov2013efficient}. Among the different versions of this technique, Continuous Bag-of-Word model and Skip-Gram model are the more commonly used \cite{goldberg2014word2vec,rong2014word2vec}.
    
    \item \textbf{FastText:} technique developed by Facebook \cite{bojanowski2017enriching} that works similarly than Word2Vec skip-gram, but overcoming two limitations of this model: it allows to incorporate subwords in the embedding process, and therefore allows to include words not contained on the original test lexicon \cite{schmitt2018joint}.
    
    \item \textbf{GloVe:} standing for Global Vectors for Word Representation, this technique was developed in Stanford \cite{pennington2014glove}, and relies on the use of a word co-occurrence matrix on which factorization techniques are applied to extract the vectors associated with each word. While Word2Vec appears to have a better performance than this technique, Glove has the advantage of having more available trained models to work with \cite{mikolov2017advances}.
    
    \end{itemize}

Analysing the application of these approaches in the reviewed articles on extremism, three different purposes can be identified: bias analysis (how pejorative terms are related to some entities and not to others) \cite{ottoni2018analyzing}, to check how two texts use similar tokens but with different meanings \cite{kursuncu2019modeling,gomes2017profiling}, or to create new lexicons based on an already checked text \cite{araque2020approach,nouh2019understanding}. Another advantage of these techniques, beyond the variety of applications they have, is that they can be used to overcome language limitations on extremist detection \cite{johnston2017identifying}.

Regarding the frequency of use of these techniques in the field of extremism, as shown in the Table \ref{tab:NLP_technique_Summary}, the technique most used (Word2Vec) does not reach 10\%, a value much lower than most of the classical techniques based on vector space models. This is due to the fact that this type of approach is becoming of great importance just in the last few years, and it is at the current time when the extension of its application in the field of extremism is taking place.

Only one article reported a comparison among FastText, Word2Vec and GloVe on an extremism classification task. FastText performed slightly better than the others two. However, Word2Vec and its variations (doc2vec, graph2vec, etc.) still remain as most used word embedding technique. Table \ref{tab:comparison-word-embedding} summarizes the comparative of these techniques in the context of extremism research.

A brief summary of the advantages and disadvantages of all these techniques appear on table \ref{tab:comparison-word-embedding}.

\begin{table}[h]
\centering
\caption{Comparison of Neural Models based techniques to generate features used in articles reviewed.}
\begin{tabular}{|l|p{6cm}|p{6cm}|}
\hline
\textbf{Technique} &
  \textbf{Advantages} &
  \textbf{Disadvantages} \\ \hline
Word2Vec &
  -Allows to predict words depending on the context. &
  -Does not recognize words not included on the trained lexicon (problematic in multilingual approaches). \\ \hline
\multirow{2}{*}{FastText} &
  -Allows to incorporate words not contained on trained lexicon. &  -Few applied in the area  \\ \hline
GloVe &
  -High amount of trained models to work with. & -Few applied in the area\\ \hline
\end{tabular}%
\label{tab:comparison-word-embedding}
\end{table}

\subsubsection{Syntactic and Semantic Features}
\label{sytactic_semantic}

There are NLP techniques based on the analysis of data according to a particular context for generating features representing the text \cite{krippendorff2018content}. The type of contextual information assessed depends on the NLP technique applied, but common approaches include sentiment analysis, topic detection or semantic analysis, among others. Techniques of this type used by the reviewed articles include:

\begin{itemize}

    \item \textbf{Part-of-Speech (POS):} allows tagging every word with its grammatical category (e.g. nouns, verbs or adjectives) depending on the structure of the text where it is found \cite{cutting1992practical}. 
    
    \item \textbf{Lexical Syntactic Feature-based (LSF):} allows capturing the dependence inside a sentence or a text between two terms \cite{benito2019design}. These two terms are later compared to determine the context and the direction of the expression. 
    
    \item \textbf{Named Entity Recognition (NER):} deals with the identification of entities (e.g. names, organizations or locations) in the text, tagging them as relevant subjects \cite{ritter2011named}.
     
    \item \textbf{Parse trees (PT):} allows to construct a representation of how the concepts can be used recursively in a sentence. Parse trees include all the tokens and their relationships, along with a set of rules that allow to substitute the token while maintaining the syntactic rules.
    
    \item \textbf{Latent Dirichlet Allocation (LDA):} is one of the most popular topic detection techniques on Natural Language Processing. It extracts topics from a corpus of text based on word probabilities: for each latent topic, it extracts the probability distribution of a combination of words, which helps to identify the main topics. \cite{jelodar2019latent}
    
    \item \textbf{Non-Negative Matrix Factorization (NMF):} is a topic modeling technique which relies on the use of linear algebra algorithms in a TF-IDF document matrix to define topics \cite{chen2019experimental}.
     
    \item \textbf{Sentiment Scoring: (SS)} provides a score for every text unit (e.g. sentence or text) based on its latent emotional valence, with the aim of understanding the authors opinion or emotional state about something \cite{feldman2013techniques}. This score can be computed as dimensional (through a single scoring about the valence) or categorical (specifying which emotions are expressed in the text). Table \ref{tab:sentiment_type} summarizes how both approaches are distributed among the reviewed articles:
    
\begin{table}[h]
\caption{Type of sentiment analysis approaches using in the reviewed articles on extremist}
\centering
\begin{tabular}{|l|l|p{7cm}|}
\hline
\textbf{Sentiment analysis approach} & \textbf{Percentage Use} &  \textbf{Articles using it} \\ \hline
Sentiment scoring (dimensional) & 32.81\% &
  \cite{wignell2018natural,owoeye2018classification,scrivens2015sentiment,hall2020machines,chen2008sentiment,macnair2018changes,figea2016measuring,scrivens2016sentiment,scrivens2018searching,weir2016positing,owoeye2019classification,mirani2016sentiment,rowe2016mining,dillon2020comparison,torregrosa2020analyzing,scrivens2020measuring,wei2016identification,bermingham2009combining,masood15using,saif2017semantic,ahmad2019detection} \\ \hline
Emotion scoring (categorical) & 9.37\% &
  \cite{wignell2018natural,chen2008sentiment,heidarysafa2020women,araque2020approach,hartung2017identifying,ahmad2019detection} \\ \hline
\end{tabular}%
\label{tab:sentiment_type}
\end{table}
    
    \item \textbf{Semantic tagging (ST):} implies the process of automatic extracting concepts, entities or topics from the  tokens in a text \cite{jovanovic2014automated}. 
    
    \item \textbf{Word/sentence length:} analyses the length of the words (based on characters) and/or the sentences (based on words) \cite{stankov2010contemporary,yang2011social,sikos2014authorship,weir2016positing,scrivens2018searching}.
    
    \item \textbf{Use of emoticons:} emoticons are built as graphical figures to express emotions or behaviours on the text, using a combination of characters \cite{agarwal2015using,wei2016identification}.
    
    \item \textbf{Use of punctuation:} this approach implies the analysis of the use of punctuation signs as part of the syntactic distribution of the sentence \cite{sikos2014authorship,yang2011social}.
    
    \end{itemize}

These types of techniques go a step further into the representation of texts, taking advantage of the tokens to conduct more complex analysis. This is specially useful in a field such as extremism research, on which simply token use or frequency can be misleading in the interpretation of outcomes \cite{fernandez2018contextual}. 

The first four techniques mentioned, POS, NER, LSF and PT, are used to analyze, tag and extract information about the syntactical structure underlying tokens. POS and NER are used to identify the nouns and entities present on the text. Then this information is used to determine which nouns from the text are actual people, organizations or locations \cite{hartung2017identifying,saif2017semantic,saif2016role,fernandez2018contextual,bisgin2019analyzing}, among others. In particular, according to the articles reviewed, NER technique shows that using a combination of noun semantic categories was statistically more accurate to determine if a text included extremist content than using token analysis, sentiment or topic features \cite{saif2017semantic,saif2016role}. Analyzing the frequency of application shown in the Table \ref{tab:NLP_technique_Summary} of these 4 techniques in the field of extremism, it can be noticed that the technique most commonly used is POS with 25\%, being the rest of the techniques very few used in comparison.

On the other hand, LSF and PT take into consideration the syntax and the dependencies among tokens. In this case, LSF analyze the relationship between two syntactically dependent tokens \cite{kim2017empirical,masood15using}, while Parse trees build representations of several tokens and use their syntactic structure to find tokens combined in the same way \cite{sikos2014authorship}. LSF was compared with Vectorial Space Models as classification feature, but it did not perform better than the latter \cite{hartung2017identifying}. 

Concerning topic extraction, LDA and NMF have been the mainly used techniques on the reviewed articles. LDA has the advantage of relying on a statistical base and to be commonly used in the literature \cite{heidarysafa2020women}. However, as a study states \cite{alizadeh2019psychology}, it performs poorly with short texts (e.g. tweets). Taking into account that most of the articles reviewed use Twitter to extract their extremist datasets, this is an important disadvantage. NMF appears as an alternative to LDA, as it appears to present more readily interpretable results \cite{o2012analysis,o2015down}, and also to have a better performance on short texts \cite{chen2019experimental}, although in the articles reviewed it is used much less frequently (see Table \ref{tab:NLP_technique_Summary}). 

Adding a topic an ``emotional value" can help forming a representative idea about the author's agreement with that topic \cite{bermingham2009combining,scrivens2018searching}. For example, two studies focused on Arabic regular population found that Twitter users tone was more negative when ISIS conducted a murdered, won a battle or made a public call or movement \cite{mirani2016sentiment,ceron2019isis}. Sentiment scoring techniques are divided in two different approaches: a dimensional approach, based on a single score, and a categorical approach, based on the classification of tokens inside one or more emotions (such as anger, fear or happiness). A combination of both strategies can be found on some of the articles reviewed \cite{wignell2018natural,figea2016measuring}. These techniques can be used to measure the emotions expressed on the text, together with the opinion of the writer towards a specific token in the text\cite{bakshi2016opinion}. The main difference among them are their theoretical approach, but also how they are applied: dimensional scoring usually involves selecting a token, around which the scoring process takes part. On the other hand, categorical scoring usually classifies tokens depending on the emotion they represent, and therefore are more focused on single tokens. In the case of extremism research, both approaches can be useful, as they can be used to identify how do extremist texts approach different topics \cite{wignell2018natural,macnair2018changes}, which valence have their tones \cite{wei2016identification} or which connotations have the terms they use \cite{chen2008sentiment}. Finally, the concept of semantic tagging was used on the articles reviewed to tag tokens with semantic information regarding their context. This strategy, very similar to NER (sometimes using it), tags the tokens with entities, but also with concepts and categories \cite{wignell2018natural}. Focusing on the use of this type of techniques in the reviewed articles, Table \ref{tab:NLP_technique_Summary} shows that the sentiment analysis techniques are the most used within the techniques to extract syntactic and semantic features, exceeding 37\% in the case of sentiment scoring. 

Last three techniques are focused on the analysis of the text formatting characteristics, to build other types of features that capture more information than that provided by the text itself. For example, the length and quantity of texts, sentences or words, the number of characters inside a word, the use of punctuation or emoticons, etc. In all the cases, text characteristic features have been used as a complement to other text features, never as single feature extracted from the free text. However, they have showed little impact to describe or predict extremism on texts, and in general all of them are being applied in few of the reviewed works (as can be seen in the last 3 rows of the  Table \ref{tab:NLP_technique_Summary}).

Table \ref{tab:comparison_content_analysis} presents a summary of all the techniques used to generate syntactic and semantic features showing their advantages and disadvantages both in general application and in extremism literature.

\begin{table}[h]
\centering
\caption{Comparison of Syntactic and Semantic based techniques to generate features for text representation.}
\begin{tabular}{|l|p{6cm}|p{6cm}|}
\hline
\multicolumn{1}{|c|}{\textbf{Technique}} &
  \multicolumn{1}{c|}{\textbf{Advantages}} &
  \multicolumn{1}{c|}{\textbf{Disadvantages}} \\ \hline
\multirow{2}{*}{POS} &
  -Allows to detect the grammatical type of tokens &
  -Regarding nouns, not as informative as NER. \\ 
 &
  -Widely used in the area with different applications (term disambiguation or classification)&
   \\ \hline
NER &
  -Detects entities, categorizing them. Useful to identify the main actors in an extremist discourse. &
  -Not as extended as POS, limited to nouns and to a trained lexicon. \\ \hline
LSF &
  -Provides a meaningful relationship among tokens. &
  -Does not perform better in the applications within the area than more simple features. \\ \hline
PT &
  -Finds sentences with an structure grammatically similar. &
  -Does not inform about the tokens itself. Not commonly used on extremism literature. \\ \hline
\multirow{2}{*}{LDA} &
  -Widely used on extremism research. &
  -Performs poorly on short texts, such as tweets (very used to conduct extremism analysis). \\
 &
  -Performs closer to a human topic classifier than other techniques. &
  -Tends to over-generalize topics \\ \hline
NMF &
  -Alternative for LDA showing a good performance on short texts. &
  -Not commonly used by authors, who tend to use LDA. \\ \hline
\multirow{2}{*}{SS (Dim.)} &
  -Simple way of measuring a sentence emotional value. &
  -Does not provide elaborate information about emotions in the sentence. \\  
 &
  -Useful to detect opinions, specially useful when combined with the detection of entities in the radical discourse. &
   \\ \hline
SC. (Cat.) &
  -Provides information about emotions in the sentence, tagging tokens and sentences with emotional categories (Happiness, sadness, anger...) &
  -Not so useful to detect opinions or tone towards a token. \\ \hline
ST &
  -As an evolution of NER, this approach "tags" nouns with their entity, concept and category. &
  -Useful to discriminate a word thanks to its context, very useful on extremism research. \\ \hline
 
  Text formatting &
  -Captures more information than those provided by the text itself. &
  -Has to be used as a complement to other text features. \\ \hline
\end{tabular}%
\label{tab:comparison_content_analysis}
\end{table}

\section{Applications of NLP in extremism research}
\label{Aplications}

Previous section has detailed all the NLP techniques used in the reviewed works on extremism to process data in text form and generate features as structured data. Depending on the objectives to be achieved in each of the reviewed works, one or several of these generated features are used to acquire new knowledge. But, in general, two main purposes have been identified in the reviewed papers for which they are used:

\begin{itemize}
    \item As a feature for classification models generated with machine learning algorithms to discriminate between extremist and non-extremist content.
    \item To conduct a descriptive analysis characterizing the extremism: for example, to detect specific slang in the case of extremism.
\end{itemize}

Based on these two main approaches, next subsections present a descriptive and comparative analysis of the works that apply each one, highlighting their outcomes.

\subsection{Classification approaches}
\label{Machine}

\begin{figure}[h]
    \centering
    \includegraphics[width=0.9\linewidth]{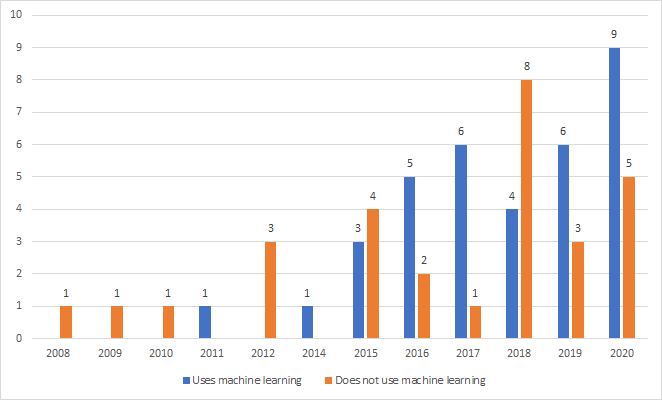}
    \caption{Frequency of articles using classification techniques vs those not using them.}
    \label{fig:methodological}
\end{figure}

As can be derived from the general analysis of the reviewed articles presented in Section \ref{general}, classification was one of the main topics of interest regarding NLP applications on extremism. This is unsurprising, as one of the key objectives of this research field is to help Law Enforcement Agencies to identify extremist content. More than half of the articles included on the review (54.68\% of the articles) applied one or more classification algorithms, specially during the first years of ISIS activity. As shown in Fig. \ref{fig:methodological}, 2015 and 2018 were the only years after the beginning of ISIS activity on which there are more articles not using classification techniques than articles using them. The common use of classification approaches shows that there was a bigger interest on detecting extremism than on defining it.

\begin{figure}[h]
    \centering
    \includegraphics[width=0.9\linewidth]{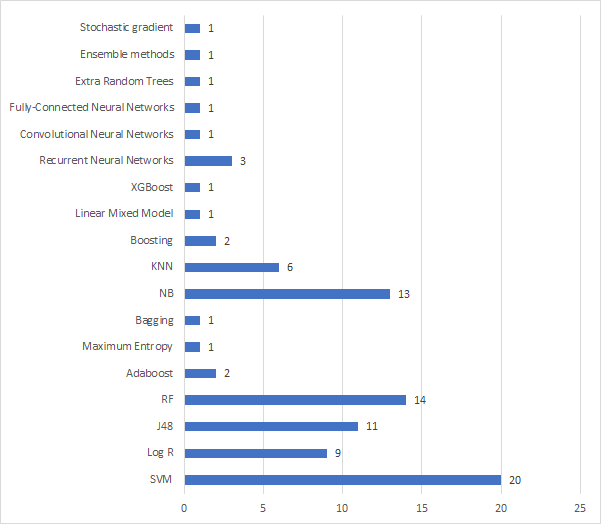}
    \caption{Type of Machine Learning algorithm used among the articles}
    \label{fig:machine_learning}
\end{figure}

With the objective of training classification models based on NLP features to discriminate between extremist and non-extremist content, different Machine Learning (ML) algorithms have been applied in the reviewed works. These works uses ML approaches to address issues that goes from sentiment tagging (using a pre-labelled dataset) to proper user classification (extremist vs non-extremist). Fig. \ref{fig:machine_learning} shows the frequency of application of every ML algorithm found on the articles reviewed, where it can be seen that Support Vector Machine (SVM) is the most commonly used model, followed by Random Forest, Naïve Bayes and Decision Tree (J48). 

Concerning the model used by each article, Table \ref{tab:ML_algorithm} summarizes what kind of Machine Learning algorithms were used by all the articles including classification tasks. It also highlights the NLP features that are directly (or indirectly) involved on the generation of the classification models. 

Apart from these classification tasks, five articles conducted other predictive learning tasks. These include the prediction of how the radicalization process takes place \cite{fernandez2018understanding}, how extremist behavioral changes occur among the members of a group \cite{smith2020detecting}, the daily level of online recruitment activities conducted by extremist groups \cite{scanlon2015forecasting}, the risk of a video to be raided by extremist groups \cite{mariconti2019you} or the risk of pro-ISIS terms as part of a person's vocabulary \cite{rowe2016mining}.

\begin{table}[!h]
\vspace{-0.1cm}
\caption{Type of features input to the ML models employed in the reviewed articles (SVM: Support Vector Machine, KNN: K-Nearest Neighbors, NB: Naïve Bayes, RF: Random Forest, Log R: Logistic Regression, LMM: Linear Mixed Models, RNN: Recurrent Neural Networks, CNN: Convolutional Neural Networks, FCNN: Fully-Connected Neural Networks, SGD: Stochastic gradient descent).}
\vspace{-0.1cm}
\footnotesize
\resizebox{0.9\columnwidth}{!}{\begin{tabularx}{\textwidth}{@{}X@{\hphantom{0}}X@{\hphantom{1}}X@{\hphantom{1}}X@{\hphantom{1}}X@{\hphantom{1}}X@{\hphantom{1}}l@{\hphantom{1}}X@{\hphantom{1}}l@{\hphantom{1}}X@{\hphantom{1}}X@{\hphantom{1}}X@{\hphantom{1}}l@{\hphantom{1}}l@{}}
\toprule
\multirow{2}{*}{\textbf{\shortstack[l]{ML\\method}}} &
  \multicolumn{13}{c}{\textbf{Features}} \\ 
 &  N-grams &  Dic. &  TF-IDF &  TF &  POS &
  NER &
  LSF &
  PT &
  SS &
  LDA &
  Emb. &
  ST &
  Others \\ \midrule
\multirow{5}{*}{SVM} &
  \cite{hartung2017identifying,masood15using,saif2017semantic,rehman2021understanding,sharif2019empirical,abddetecting2020} &
  \cite{figea2016measuring,sikos2014authorship,yang2011social,agarwal2015using,rehman2021understanding} &
  \cite{yang2011social,rehman2021understanding,sharif2019empirical,masood15using,kim2017empirical,abddetecting2020} &
  \cite{hartung2017identifying,masood15using,scanlon2015forecasting,ahmad2019detection,devyatkin2017exploring,kim2017empirical,mirani2016sentiment,chen2008sentiment,figea2016measuring,wei2016identification,agarwal2015using,araque2020approach,fernandez2018contextual} &
  \cite{figea2016measuring,sikos2014authorship,yang2011social,devyatkin2017exploring} &
  \cite{hartung2017identifying,yang2011social} &
  \cite{hartung2017identifying,masood15using,kim2017empirical} &
  \cite{sikos2014authorship} &
  \cite{figea2016measuring,mirani2016sentiment,wei2016identification,masood15using,saif2017semantic,yang2011social,ahmad2019detection,araque2020approach,hartung2017identifying} &
  \cite{saif2017semantic,scanlon2015forecasting,kim2017empirical} &
  \cite{araque2020approach,masood15using,devyatkin2017exploring,kim2017empirical,abddetecting2020} &
  \cite{saif2017semantic,fernandez2018contextual,devyatkin2017exploring} &
  \cite{sikos2014authorship,yang2011social} \\ \midrule
\multirow{2}{*}{KNN} &
  \cite{sharif2019empirical,abddetecting2020} &
  \cite{agarwal2015using} &
   
  \cite{sharif2019empirical,sharif2020detecting,abddetecting2020} &
  \cite{ahmad2019detection,wei2016identification,agarwal2015using} &
  &
   &
   &
   &
  \cite{wei2016identification,ahmad2019detection} &
   &
   &
   &
   \\ \midrule
\multirow{5}{*}{NB} &
  \cite{masood15using,rehman2021understanding,sharif2019empirical,sharif2020detecting,abddetecting2020} &
  \cite{yang2011social,rehman2021understanding,fernandez2018understanding} &
  \cite{yang2011social,zahra2018framework,rehman2021understanding,sharif2019empirical,masood15using,abddetecting2020} &
  \cite{masood15using,scanlon2015forecasting,ahmad2019detection,saif2016role,devyatkin2017exploring,sharif2020detecting,wei2016identification,fernandez2018understanding,kursuncu2019modeling,fernandez2018contextual} &
  \cite{yang2011social,devyatkin2017exploring} &
  \cite{yang2011social} &
  \cite{masood15using} &
   &
  \cite{wei2016identification,masood15using,yang2011social,ahmad2019detection} &
  \cite{scanlon2015forecasting} &
  \cite{masood15using,devyatkin2017exploring,kursuncu2019modeling,abddetecting2020} &
  \cite{saif2016role,fernandez2018contextual,devyatkin2017exploring} &
  \cite{yang2011social} \\ \midrule
\multirow{2}{*}{Boosting} &
   &
   &
   &
  \cite{scanlon2015forecasting,devyatkin2017exploring}
  \cite{devyatkin2017exploring} &
  &
   &
   &
   &
   &
  \cite{scanlon2015forecasting} &
  \cite{devyatkin2017exploring} &
  \cite{devyatkin2017exploring} &
   \\ \midrule
\multirow{4}{*}{J48} &
  \cite{sharif2019empirical,rekik2020recursive,sharif2020detecting,abddetecting2020} &
  \cite{fernandez2018understanding} &
  \cite{sharif2019empirical,sharif2020detecting,masood15using,abddetecting2020} &
  \cite{sharif2020detecting,mirani2016sentiment,owoeye2019classification,rekik2020recursive,owoeye2018classification,fernandez2018understanding,fernandez2018contextual} &
  \cite{owoeye2018classification} &
   &
   &
   &
  \cite{owoeye2018classification,scrivens2016sentiment,weir2016positing,owoeye2019classification,mirani2016sentiment} &
   &
  \cite{abddetecting2020} &
  \cite{fernandez2018contextual} &
  \cite{weir2016positing} \\ \midrule
\multirow{5}{*}{RF} &
  \cite{masood15using,de2020radical,rehman2021understanding,sharif2019empirical,sharif2020detecting,abddetecting2020,nouh2019understanding} &
  \cite{figea2016measuring,rehman2021understanding,nouh2019understanding} &
  \cite{ahmad2019detection,mariconti2019you,rehman2021understanding,sharif2019empirical,sharif2020detecting,abddetecting2020,nouh2019understanding} &
  \cite{masood15using,mariconti2019you,ahmad2019detection,devyatkin2017exploring,sharif2020detecting,mirani2016sentiment,figea2016measuring,de2020radical,kursuncu2019modeling} &
  \cite{figea2016measuring,devyatkin2017exploring,de2020radical} &
   &
  \cite{masood15using} &
   &
  \cite{figea2016measuring,weir2016positing,mirani2016sentiment,masood15using,ahmad2019detection,nouh2019understanding} &
   &
  \cite{masood15using,devyatkin2017exploring,kursuncu2019modeling,abddetecting2020,nouh2019understanding} &
  \cite{devyatkin2017exploring} &
  \cite{weir2016positing,de2020radical} \\ \midrule
Adaboost &
   &
  \cite{figea2016measuring,yang2011social} &
  \cite{yang2011social} &
  \cite{figea2016measuring} &
  \cite{figea2016measuring,yang2011social} &
  \cite{yang2011social} &
   &
   &
  \cite{figea2016measuring,yang2011social} &
   &
   &
   &
  \cite{yang2011social} \\ \midrule
\multirow{4}{*}{Log R} &
  \cite{masood15using,sharif2020detecting,abddetecting2020} &
  \cite{smith2020detecting,fernandez2018understanding} &
  \cite{sharif2020detecting,masood15using,abddetecting2020} &
  \cite{masood15using,devyatkin2017exploring,sharif2020detecting,wei2016identification,smith2020detecting,fernandez2018understanding,araque2020approach} &
  \cite{devyatkin2017exploring} &
   &
  \cite{masood15using} &
   &
  \cite{wei2016identification,masood15using,araque2020approach} &
   &
  \cite{araque2020approach,masood15using,johnston2020identifying,devyatkin2017exploring,abddetecting2020} &
  \cite{devyatkin2017exploring} &
   \\ \midrule
LMM &
   &
  \cite{smith2020detecting} &
   &
  \cite{smith2020detecting} &
  &
   &
   &
   &
   &
   &
   &
   &
   \\ \midrule
XGBoost &
   &
   &
   \cite{kim2017empirical} &
   &
   &
   &
  \cite{kim2017empirical} &
   &
   &
  \cite{kim2017empirical} &
  \cite{kim2017empirical} &
   &
   \\ \midrule
Maximum Entropy &
   &
   &
   &
  \cite{mirani2016sentiment} &
  &
   &
   &
   &
  \cite{mirani2016sentiment} &
   &
   &
   &
   \\ \midrule
Bagging &
   &
   &
   &
  \cite{mirani2016sentiment} &
  &
   &
   &
   &
  \cite{mirani2016sentiment} &
   &
   &
   &
   \\ \midrule
RNN &
   &
   &
   &
  \cite{mariconti2019you} &
  &
   &
   &
   &
  \cite{ahmad2019detection} &
   &
  \cite{johnston2020identifying,ahmad2019detection} &
   &
   \\ \midrule
CNN &
   &
   &
   &
   &
   &
   &
   &
   &
  \cite{ahmad2019detection} &
   &
  \cite{ahmad2019detection} &
   &
   \\ \midrule
FCNN &
   &
   &
   &
   &
   &
   &
   &
   &
   &
   &
  \cite{johnston2017identifying} &
   &
   \\ \midrule
Extra Random Trees &
   &
   &
   \cite{mariconti2019you} &
  \cite{mariconti2019you} &
  &
   &
   &
   &
   &
   &
   &
   &
   \\ \midrule
Ensemble methods &
  \cite{sharif2019empirical} &
   &
   \cite{sharif2019empirical} &
   &
   &
   &
   &
   &
   &
   &
   &
   &
   \\ \midrule
SGD &
  \cite{sharif2020detecting} &
   &
   \cite{sharif2020detecting} &
  \cite{sharif2020detecting} &
  &
   &
   &
   &
   &
   &
   &
   &
   \\ \midrule
\end{tabularx}}%
\vspace{-0.1cm}
\label{tab:ML_algorithm}
\end{table}

Focusing on the use of basic features based on vectorial space models, such as n-grams and dictionaries (shown in Table \ref{tab:ML_algorithm}), the first ones \cite{bisgin2019analyzing,hartung2017identifying,kursuncu2019modeling,owoeye2018classification,rekik2019violent,scanlon2015forecasting,sharif2019empirical,zahra2018framework} has been used more than the second ones \cite{ahmad2019detection,araque2020approach,fernandez2018understanding,kursuncu2019modeling}.
It would be difficult to determine which of these two techniques performs better. In fact, the study of Figea et al. \cite{figea2016measuring} found that there is no relevant difference between using dependent techniques (such n-grams) or independent (such as LIWC) from the text when creating a classification model. A general limitation from both techniques is that similar terms can be used with different meanings in two texts, leading to confusions on the data interpretation process \cite{saif2016role,fernandez2018contextual,wei2018detecting,gomes2017profiling}. This is common in the context of religious radicalization, where religious terms can be used by regular religious texts, but also by extremists texts \cite{gomes2017profiling}. While using n-grams (when $n > 1$) is a way to overcome this limitation, they are a primitive option to keep semantic information  \cite{hall2020machines,sharif2019empirical}. There are, however, techniques that are more informative than these to conduct complex NLP analysis. For example, n-grams were found to be less able to identify topics in radical texts than LDA or dictionaries \cite{hall2020machines}.

Regarding sentiment features, they are not usually used as a single feature to detect extremist content, specially concerning political radicalisation \cite{scrivens2015sentiment}. While the type of features do not perform bad either and they, in fact, perfform better than other less complex features \cite{ahmad2019detection}, usually classification models trained with more features perform better than those who use only sentiment features \cite{weir2016positing,hartung2017identifying,saif2017semantic,owoeye2018classification,owoeye2019classification,araque2020approach}. In fact, those classifiers based on semantic features exclusively performed better than those based on sentiment features exclusively \cite{saif2017semantic,araque2020approach}. For example, a study conducted by Weir et al. \cite{weir2016positing} compared the usefulness of two classification tools, one based on sentiment features and the other using POS feature together with text formatting features such as number of sentences, average length or quantity of characters. The second showed a better performance, but it could be due to the high number of features used on it. Other three articles \cite{sikos2014authorship,yang2011social,stankov2010contemporary} also used text formatting features and other text features, as models to describe and classify extremist content. None showed a significant difference from classifiers that only use features that extract information from the text itself. But there are several works which conclude that text text formatting features (such as sentence length\cite{yang2011social} or emoticons \cite{agarwal2015using,wei2016identification}) are a good add-on to improve the accuracy of the classification models.


Finally, the best classification outcomes are achieved using features based on Neural Language Models (word embedding). Articles using this type of textual representation as classification feature found that it tends to perform better than other classical features such as vectorial space models \cite{devyatkin2017exploring,kursuncu2019modeling,masood15using}, or syntactic and semantic features \cite{kim2017empirical,araque2020approach}. One article, however, pointed that word embedding tend to perform poorly than n-grams on short pieces of text \cite{abddetecting2020}. As happened with other NLP features, it was found that combining word embedding based features with the other types also showed better classification outcomes than using it isolated \cite{araque2020approach,nouh2019understanding}. 

The main purpose of most articles that use features based on Neural Language Models in classification tasks is the detection of extremist content. As other types of features, they are quite dependent on the type of machine learning algorithm used \cite{masood15using,kim2017empirical,johnston2020identifying,devyatkin2017exploring}, but they work specially good when combined with neural networks of different type \cite{ahmad2019detection,johnston2017identifying}. They are also good to detect radical users, but have been found to perform poorer than n-grams to detect extremism on small pieces of text \cite{abddetecting2020}.

\subsection{Descriptive approaches}
\label{Descriptive}

A second application of NLP techniques in extremism research found is the characterization and study of the phenomenon of extremism from a descriptive perspective. Within these works, 5 different descriptive focus can be identified:

\begin{itemize}
    \item \textit{Terms}: descriptive analysis on the terms commonly used by extremists. Characterization of the type of extremist vocabulary.
    \item \textit{Topics}: detection of the most common topics discussed by extremist texts.
    \item \textit{Sentiment}: analysis of the sentiment and tone of an extremist discourse.
    \item \textit{Semantic}: analysis of the contextual information around terms inside an extremist text.
    \item \textit{Punctuation}: descriptive analysis of the text format commonly used in the extremist environment.
\end{itemize}

Table \ref{tab:descriptive_approach} summarizes the type of descriptive analysis performed for each of the articles reviewed. The most simple descriptive approach would focus on the terms, while the inclusion of other approaches (topics, sentiment, semantic or punctuation) add extra layers to the description of the discourse. This is why the terms approach is the most common. In addition, we can see that almost all the rest of descriptive analyses have previously performed a term analysis, showing that all the approaches are complementary. Sentiment analysis is the only one that is occasionally performed independently.

\begin{table}[h]
\centering
\caption{Descriptive linguistic approach used by the reviewed articles.}
\begin{tabular}{|l|l|p{8cm}|}
\hline
\textbf{Descriptive linguistic approach} & \textbf{Percentage Use}  &
  \textbf{Articles using it} \\ \hline
Terms & 67.85\% &
  \cite{heidarysafa2020women,rekik2019violent,kinney2018theming,gomes2017profiling,torregrosa2020analyzing,alizadeh2019psychology,bisgin2019analyzing,hall2020machines,stankov2010contemporary,prentice2012language,ben2016hate,alghamdi2012topic,bermingham2009combining,klein2019online,abdelzaher2019systematic,wei2018detecting,wignell2018natural,macnair2018changes,skillicorn2015empirical} \\ \hline
Topics & 46,42\% &
  \cite{heidarysafa2020women,kinney2018theming,alizadeh2019psychology,bisgin2019analyzing,hall2020machines,ben2016hate,alghamdi2012topic,bermingham2009combining,klein2019online,o2012analysis,ottoni2018analyzing,o2015down,wadhwa2015approach} \\ 
  \hline
Sentiment & 39.28\% &
\cite{heidarysafa2020women,bermingham2009combining,torregrosa2020analyzing,wignell2018natural,macnair2018changes,chen2008sentiment,scrivens2020measuring,dillon2020comparison,scrivens2018searching,scrivens2015sentiment,alizadeh2019psychology} \\ 
  \hline
Semantic & 17.85\% &
\cite{wignell2018natural,ottoni2018analyzing,gomes2017profiling,prentice2012language,abdelzaher2019systematic} \\ 
  \hline
Punctuation & 3.57\% &
  \cite{stankov2010contemporary} \\ \hline
\end{tabular}%
\label{tab:descriptive_approach}
\end{table}

Regarding the insights about extremism found in the reviewed works, sections \ref{insight_religious} and \ref{insight_farright} highlight the main patterns observed, classified by the two predominant types of extremism found in Section \ref{general}: Religious and Political. Table \ref{tab:comparison-extremist-discourse} introduces a summary and comparison between the two most studied extremist movements, Jihadism and Far-right, which are explained in detail below.

\begin{table}[h]
\centering
\caption{Comparison of discourse insights from the most commonly mentioned extremist groups' discourse.}
\begin{tabular}{|p{1.5cm}|p{6.3cm}|p{6.3cm}|}
\hline
\textbf{Insights} &
  \textbf{Jihadi extremism} &
  \textbf{Far-right extremism} \\ \hline
\multirow{2}{*}{Terms} &
  -Religious terms, geographical references. &
  -Supremacist, racist, antiimmigration and anti-left terms. Specific slang regarding these \\ 
 &
  -Specific slang related to the religious conflict (e. g. ``Crusaders", ``Kaffir", etc.). &
  -Specific slang regarding the previously mentioned terms (e.g. ``Illegal Aliens", ``WhiteGenocide", ``14-88", etc.) \\ \hline

Topics &
  -Religion, war, geopolitics, extremist philosophy, recruitment, military. &
  -Politics, racial topics, immigration and war. \\ \hline
\multirow{4}{*}{Sentiment} &
  -Jihadi women tend to be more extreme on their messages than men. &
  -Negative messages directed against Jews, LGBT and black people. \\ 
 &
  -General presence of negative tone. &
  -General presence of negative tone \\ 
 &
  -Words related with emotions of fear, hate and violence, except when talking about topics such as paradise or martyrdom. &
  -Use of anger, disgust and negativity related terms. \\ 
 &
  -A positive tone towards ISIS can be related to a complicity with this group. &
   \\ \hline
Semantic &
  -Preference for terms such as ``Islamic State" or ``Caliphate", instead of ISIS. Entities are a good way to discriminate between a regular or an extremist use of a term. &
  -Common semantic categories include ``violence" and ``anger" \\ \hline
  Punctuation &
  -Frequent use of Arabic terms, even in non-Arabic texts. &
   \\ \hline
\end{tabular}%
\label{tab:comparison-extremist-discourse}
\end{table}

\subsubsection{Literature insights about religious extremism}
\label{insight_religious}

Concerning common terms used by religious extremism, the name "ISIS" was more mentioned by neutral users than by extremist users \cite{wignell2018natural,gomes2017profiling,bisgin2019analyzing}, who preferred the term "Islamic State" or "Caliphate".The more frequent terms found in the extremist text analyzed in the articles were related to religious (e.g. Allah, Jihad or Islam) or geographical references (e.g. Syria, Raqqa, America or Iraq) \cite{wignell2018natural,gomes2017profiling,wei2018detecting,bisgin2019analyzing,skillicorn2015empirical}. The descriptive analysis of the text also detected the common use of specific slang terms, such as "Crusaders", "Mujahideen" or "Abu" \cite{gomes2017profiling,wei2018detecting}.  

The works carrying out an descriptive analysis focused on the topics shows that the most frequent topic related to Jihadi extremism was, unsurprisingly, religion  \cite{scanlon2015forecasting,bermingham2009combining,kinney2018theming}. Jihadi magazines more easily identifiable topics were war, geopolitics, religious speech, government and administration \cite{bisgin2019analyzing}. Inspire (Al Qaeda's magazine) was more focused on conflict legitimisation and philosophy, while Dabiq and Rumiyah (ISIS magazine) were more focused on the geopolitical conflict \cite{kinney2018theming}. Some of the topics, such as recruitment, were found hidden among topics referring to religious and military aspects of the Syria conflict\cite{scanlon2015forecasting}.

Combining sentiment analysis and topic detection, jihadi women were found more extreme than men on their messages on nearly every relevant topic  \cite{bermingham2009combining}. Concerning the journals, it was found that most of their texts had negative tone, while using terms related to fear, except when they discussed about topics such as paradise or martyrdom \cite{wignell2018natural,macnair2018changes}. Words such as Allah or Islamic State were also found to have negative connotations when analyzed through a sentiment analysis approach. Authors hypothesize that this can be due to their use use as justification of violent behaviours. A study concerning jihadi radical forums also found that the most extremist texts also scored more on negative dimensions, using violence and hate terms, than those more moderate \cite{chen2008sentiment}. Finally, a study hypothesized that radical users that presented a good tone towards ISIS (on their tweets) showed in fact complicity with it \cite{wei2016identification}.

While the descriptive term analysis approach helps providing a first insight, it shall be remembered that context can vary the meaning of a token \cite{wei2018detecting}. From this perspective, articles focused on semantic discrimination allowed to check how these keywords are used depending on the intention of the text. For example, Gomes et al. \cite{gomes2017profiling} stated that the background of the terms "ISIS", "Islamic" and "Syria" changes depending on the origin of the text analysed (neutral or extremist). A study analyzing divergences on the semantic meaning of words, conducted by Fernandez et al. \cite{fernandez2018contextual}, classified terms into different semantic groups (category, entity and type of entity). It was found that similar words were used differently by radical and non-radical users, including the name of radical groups. Entities were found to be a good way of discriminating the semantic meaning of a term. Finally, the study of Kursunku et al. \cite{kursuncu2019modeling} conducted a comparative analysis between extremist and non-extremist religious users. They found that, while both groups shared terminology when referring to the religious concept, the extremist group used much more terms related to radical Islamism and hate speech. This is why using token analysis techniques combined with other strategies can be more informative than using it isolated.

As can be stated by these insights, and taking into consideration the features of an extremist discourse presented on section \ref{operativization}, Jihadi extremism presents several of these features. Their use of specific slang and expressions, together with a negative tone, shows how they present and specific linguistic style. Also, they build their discourses with a special emphasis on a theological and moral narrative, but also with the glorification of religious acts of violence against a common enemy (Western society and non-believers). While it is difficult to determine how much their use of war topics is related with a specific narrative or the geopolitical situation of the territories on which they operated, it can also be stated that war (and its instrumentalization) its a key element on the construction of their narrative.

\subsubsection{Literature insights about political extremism}
\label{insight_farright}

Regarding the reviewed works focused on conducting a descriptive analysis of the terms most commonly used by far-right extremism, an article analyzing an Alt-right community \cite{torregrosa2020analyzing} found that they used racist (BlackMagic, WhitesLivesMatter), anti-immmigration (BuildTheWall, IllegalAliens) supremacist (WhiteGenocide, WhitePeople, ChasingDownWhites) and anti-left (AntifaTerrorists) terms and hashtags on their tweets. This work also found to use specific slang to refer to other racial minorities, such as "aliens" to refer to immigrants. Among a sample of videos massively attacked by far right groups from 4chan, some of the most mentioned keywords were ``black", ``police", ``white", ``shot", ``gun", ``world", ``war", ``American", ``government" or ``law". \cite{mariconti2019you}. Other relevant keywords on far-right extremist groups can be the mention of the numbers ``14" (a reference to the ``fourteen words", a white nationalist slogan) and ``88" (meaning ``Heil Hitler", as the H is the 8th letter of the alphabet), but also to the genocide, nazism, anti-islamic and anti-jewish groups \cite{o2012analysis,o2015down}.


Concerning the analysis of topics in political extremist groups, it was found that the more common topics discussed by far-right groups were racial topics \cite{ottoni2018analyzing,ben2016hate,alizadeh2019psychology,o2015down}, immigration \cite{ottoni2018analyzing,ben2016hate} and war \cite{ottoni2018analyzing}, being very aggressive with these topics \cite{mariconti2019you}. This was unsurprising, as both racial content, war and immigration are topics commonly found on the far-right discourse \cite{panizo2019describing}. An interesting pattern was to find that non institutional groups were more focused on a racial and anti-immigration discourse \cite{ben2016hate,klein2019online} than the institutional far-right groups, such as political parties. Those parties were occasionally found to have a populist discourse directed against the elites \cite{klein2019online}. The only article analysing far-left groups found that they discussed more about feeling related topics than other groups\cite{alizadeh2019psychology}.

Regarding sentiment analysis, one of the reviewed articles \cite{torregrosa2020analyzing} also found that a higher relevance on a far right community was related to a significantly higher use of negative and aggressive terminology. Similarly, the study of Figea et al. \cite{figea2016measuring} found that words of anger can also be useful to identify emotional concepts related to political extremist content, such as aggressiveness and concerns about other groups. Also, high negative messages were commonly directed against Jews, LGBT and black people (specially the first two) \cite{scrivens2020measuring}.


Only one article \cite{alizadeh2019psychology} focused on analyzing differences between far-right and far-left discourses, using a dictionary-based approach (both LIWC and Moral Foundation dictionaries). For these purpose the authors combine different NLP features to achieve a descriptive analysis from different perspectives, terms, topics and feelings. They found that far-right used more positive words, together with terms regarding obedience to authority and pureness, while far-left used more negative terms, anxiety words and terms related with justice and harm avoidance. Concerning a sentiment approach, this study found that both groups used a general negative tone compared to non-extremist political groups. However, from all the previously mentioned outcomes, only obedience to authority words showed a significant difference. 

Finally, the only reference to semantic analysis in political extremism related articles appear on Ottoni et al. \cite{ottoni2018analyzing}, who detected that terms from extremist groups tend to be classified in "negative" categories using the semantic tagger from Empath. Among this category, the more relevants were anger and violence. 

As it happened with religious extremism, far-right extremism also presented several features of the extremist discourses presented on section \ref{operativization}. One of their most relevant traits is their use of specific and aggressive slang to refer to other groups. However, this is not specially surprising, considering that some of these groups are very active on the Internet. They rely on political and historical narratives to build their discourse, also including a component of ``self-victimization" on it. They also use hate speech and otherness as discursive resources (specially the first one, compared to religious extremism), and frequently include references to war narrative.

\section{NLP Dataset \& Tools }
\label{Software}
In the analysis carried out in Section \ref{general}, it was noted that
the sources and the specific tool used for NLP appear frequently as relevant keywords of the articles. This is because they are a fundamental part of any research work related to the study of a particular domain, in this case the extremism phenomena. The following subsections present a detailed description of both the data sources and tools used in the works reviewed.

\subsection{Datasets and datasources}
\label{Datasets}

Obtaining a dataset is a key part of the NLP research process. In the case of online extremism, this step becomes specially difficult, as most of the information represents a risk for security or anonymity. Therefore, it becomes a hard task to find open datasets online.

\begin{table}[h]
\centering
\caption{Publicly available extremist datasets.}
\begin{tabular}{|p{3.5cm}|l|l|p{2.5cm}|p{2cm}|}
\hline
\textbf{Dataset} &
  \textbf{Size} &
  \textbf{Language} &
  \textbf{Source} &
  \textbf{Articles using this source} \\ \hline
Al-Firdaws \cite{AZSecure-AlFirdaws} &
  39.715 posts - 2.187 users &
  Arabic &
  Dark web forum &
  \cite{chen2008sentiment} \\ \hline
Montada \cite{AZSecure-Montada} &
  1.865.807 posts - 52.546 users &
  Arabic &
  Dark web forum &
  \cite{chen2008sentiment} \\ \hline
Ansar1 \cite{AZSecure-Ansar1} &
  29.492 posts - 382 users &
  English &
  Dark web forum &
  \cite{scanlon2015forecasting} \\ \hline
How ISIS uses Twitter (Kaggle) \cite{KaggleHowIsis} &
  17.410 tweets - 112 users &
  English &
  Twitter &
  \cite{araque2020approach,zahra2018framework,fernandez2018contextual,rehman2021understanding,kursuncu2019modeling,fernandez2018understanding,abddetecting2020,nouh2019understanding,gomes2017profiling} \\ \hline
Automated Hate Speech Detection and the   Problem of Offensive Language \cite{Davidson-dataset} &
  24.802 tweets - N/A users &
  English &
  Twitter &
  \cite{johnston2020identifying} \\ \hline
Crisis Lex Dataset (not specified) \cite{Crisis-dataset} &
  Not specified &
  English &
  Twitter &
  \cite{zahra2018framework} \\ \hline
UDI-TwitterCrawl-Aug2012 \cite{UDI-dataset2012} &
  50.000.000 tweets - 147.909 users &
  English &
  Twitter &
  \cite{agarwal2015using} \\ \hline
Dataset-ATM-TwitterCrawl-Aug2013 \cite{UDI-dataset2013} &
  5.000.000   tweets - N/A users &
  English &
  Twitter &
  \cite{agarwal2015using} \\ \hline
Religious Texts Used By ISIS \cite{KaggleReligiousTexts} &
  2,685 religious texts &
  English &
  Religious texts 
  &
  \cite{rehman2021understanding,abddetecting2020} \\ \hline
Tweets targeting ISIS \cite{KaggleTargeting} &
  122.000 tweets - 95.725 users &
  English &
  Twitter &
  \cite{rehman2021understanding,abddetecting2020,nouh2019understanding} \\ \hline
Gawaher \cite{AZSecure-Gawaher} &
  372.499 posts - 9.629 users &
  English & Dark web forum &
  \cite{scrivens2015sentiment,scrivens2018searching} \\ \hline
Turn to Islam \cite{AZSecure-Turn} &
  335.338 posts - 10.858 users &
  English &
  Dark web forum &
  \cite{scrivens2015sentiment,scrivens2018searching} \\ \hline
\end{tabular}%
\label{tab:datasets}
\end{table}

Much of the articles included on the review use their owns datasets. The reader is encouraged to contact with the authors of the different articles to ask for their data. However, this section deals with the articles which used datasets that are either public, or can be obtained from their original source. Table \ref{tab:datasets} shows a summary of the publicly available datasets used by the literature. This table contains the name of the dataset, an approximation to its size, the original language, the source of the data,the articles using those datasets and a bibliographic reference including a link to the dataset itself.

Also, there are data sources which are often used to extract texts, but that are not pre-processed in the way datasets are. Table \ref{tab:data_sources} presents the different extremist journals used by the literature to conduct NLP analysis. The data from these sources, however, shall be curated before conducting any analysis.

It shall be stated that, besides the already mentioned datasets (who are part of this review), there are other sources that might be useful for the person interested on obtaining more textual data related to the topic of extremism and radicalization. While these datasets are not used by the reviewed documents, and therefore remain outside of this article scope, the authors considered interesting to highlight some of them in order to help researchers to find more publicly available data. As with the type of extremism of the articles in this review, they will be divided into two groups: political and religious extremism. 

Concerning political extremism, a dataset of the far-right forum named Stormfront \cite{gibert2018hate} can be found on GitHub\footnote{https://github.com/Vicomtech/hate-speech-dataset}. Also, a dataset of alt-right users was validated by Thorburn et al. \cite{thorburn2018measuring}, which is publicly available under request to authors. Besides, speeches from different political parties can be found on the webpage of the Manifesto Project Database\footnote{https://manifestoproject.wzb.eu/}, with textual data from parties with different ideologies.

Finally, related to religious extremism, the Global Terrorism Research Project (which is the source to download Inspire magazine on Table \ref{tab:datasets}) present much more content than previously stated, including more magazines or datasets\footnote{http://gtrp.haverford.edu/resources/}. Same happens with the AZSecure webpage, which contains datasets from dark web jihadist forums in different languages\footnote{https://www.azsecure-data.org/dark-web-forums.html}.

\begin{table}[h]
\centering
\caption{Publicly available extremist data sources.}
\begin{tabular}{|l|l|l|}
\hline
\textbf{Data source} &
  \textbf{Type of source} &
  \textbf{\begin{tabular}[c]{@{}l@{}}Articles using this \\ source\end{tabular}} \\ \hline
Dabiq \cite{GTRP-Dabiq} &
  Extremist magazine &
  \cite{macnair2018changes,kinney2018theming,wignell2018natural,bisgin2019analyzing,araque2020approach,johnston2017identifying,johnston2020identifying,de2020radical,heidarysafa2020women,skillicorn2015empirical} \\ \hline
Rumiyah \cite{GTRP-Rumiyah} &
  Extremist magazine &
  \cite{macnair2018changes,kinney2018theming,wignell2018natural,araque2020approach,johnston2017identifying,johnston2020identifying,de2020radical,heidarysafa2020women} \\ \hline
Inspire \cite{GTRP-Inspire} &
  Extremist magazine &
  \cite{sikos2014authorship,johnston2017identifying,johnston2020identifying,skillicorn2015empirical} \\ \hline
Azan \cite{Archive-Azan} &
  Extremist magazine &
  \cite{skillicorn2015empirical} \\ \hline
\end{tabular}%
\label{tab:data_sources}
\end{table}

\subsection{Tools}
While conducting a research work, authors shall consider which tools are they using for their experiments, along with which knowledge bases do they use, for example, to create a lexicon. In this section, the more frequently used NLP tools when studying extremism and radicalisation will be reviewed.

\begin{figure}[h]
    \centering
    \includegraphics[width=0.65\linewidth]{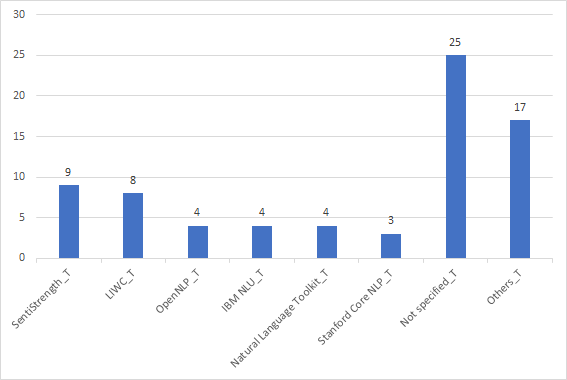}
    \caption{NLP tools used among the articles}
    \label{fig:software}
\end{figure}

Fig. \ref{fig:software} shows the frequency of use of different NLP tools. Only those being used on three or more articles have their own category, while the rest are included under the "others" category. Also, the category "non specified" includes all those articles not clearly mentioning the software tools they used \cite{chen2008sentiment,alghamdi2012topic,rowe2016mining,wei2018detecting,scanlon2015forecasting,hartung2017identifying,zahra2018framework,sharif2019empirical,fernandez2018understanding}.

The most frequently used NLP tools are:

\begin{itemize}
    \item \textbf{SentiStrength\footnote{http://sentistrength.wlv.ac.uk/}:} this tool, developed in 2010 \cite{thelwall2010sentiment}, was created to analyse the emotional valence (sentiment) of short texts. It uses a dictionary with sentiment related terms, from which it calculates the "strength" of the tone of different expressions. SentiStreght can report binary (positive vs negative), trinary (positive/negative/neutral) and single scale (-4 to +4) sentiment results. From the reviewed articles, it was the most commonly used tool to determine sentiment \cite{weir2016positing,scrivens2016sentiment,wei2016identification,saif2017semantic,owoeye2019classification,scrivens2015sentiment,macnair2018changes,scrivens2020measuring,scrivens2018searching}.
    
    \item \textbf{Linguistic Inquiry Word Count\footnote{http://liwc.wpengine.com/}:} this tool, also known as LIWC \cite{pennebaker2001linguistic}, was created on 2007 with the aim of studying the language through a psychological perspective. LIWC relies on the usage of pre-established dictionaries (which can be expanded with dictionaries made by the researcher) that are used to identify categories of words and psycho-linguistic processes underlying a text  \cite{tausczik2010psychological} . Eight articles used it to conduct their analysis \cite{alizadeh2019psychology,hall2020machines,smith2020detecting,sikos2014authorship,figea2016measuring,nouh2019understanding,torregrosa2020analyzing,rehman2021understanding}.
    
    \item \textbf{OpenNLP\footnote{https://opennlp.apache.org/}:} Apache OpenNLP library is a machine learning based toolkit for the processing of natural language text \footnote{https://opennlp.apache.org/docs/}, coded in Java. It supports different NLP tasks, providing several options to analyse texts. Four articles used OpenNLP on the review \cite{scrivens2018searching,scrivens2015sentiment,scrivens2016sentiment,weir2016positing}.
    
    \item \textbf{IBM Watson Natural Language Understanding\footnote{https://www.ibm.com/watson/natural-language-processing}:} this software, developed by IBM, includes in fact several packages inside it, which allows to conduct NLP analysis from different approaches (for example, open analysis vs questions and answers). This software can apply several NLP techniques to texts, such as semantic tagging, sentiment scoring or keywords and topic extraction. It was used by two articles on the review \cite{ahmad2019detection,wignell2018natural}. Also, the software AlchemyAPI, which was used by other two articles from the review  \cite{saif2017semantic,saif2016role}, was included in the core of Watson NLU in 2015\footnote{https://www.ibm.com/cloud/blog/announcements/bye-bye-alchemyapi}.

    \item \textbf{Natural Language Toolkit\footnote{https://www.nltk.org/}:}  the Natural Language Toolkit (NLTK) is a NLP Python library created on 2002 \cite{loper2002nltk}. It performs very similar NLP tasks than OpenNLP. Four articles used this algorithm \cite{ben2016hate,heidarysafa2020women,kinney2018theming,klein2019online}.
    
    \item \textbf{Stanford Core NLP\footnote{https://stanfordnlp.github.io/CoreNLP/}:} the Stanford CoreNLP is another Java based NLP tool, developed on Stanford \cite{manning2014stanford}. It can perform analysis in different languages, and one its main features is that it is quite easy to set up and run \cite{pinto2016comparing}. Three articles used this NLP tool \cite{wei2016identification,kim2017empirical,bisgin2019analyzing}.
    \end{itemize}
    
Even though Fig. \ref{fig:software} summarizes the most used NLP tools, other tools are used by less than three reviewed articles. These tools include WordNet \cite{bermingham2009combining}, Stanford Maximum Entropy Part-of-speech Tagger \cite{bermingham2009combining}, Vader\cite{wei2016identification,torregrosa2020analyzing}, WMatrix \cite{prentice2012language}, Gensim \cite{ottoni2018analyzing}, iSA \cite{ceron2019isis}, the Arules Package \cite{rekik2019violent}, MALLET \cite{hall2020machines}, the Language Detection Library for Java \cite{agarwal2015using}, POSIT \cite{weir2016positing,owoeye2018classification}, TextRazor \cite{fernandez2018contextual}, Language Model Toolkit  \cite{mariconti2019you}, ConcepNet \cite{mariconti2019you}, TensorFlow Vocabulary Processor \cite{johnston2020identifying} and the Python-based tone analyzer API \cite{ahmad2019detection}.

\section{Discussion and Conclusion}
\label{Discussion}

This review aimed to explain the contributions that NLP has provided for extremism research so far. This  interest was divided in several research questions presented in the introduction, regarding the different NLP issues analyzed. Through the whole article, those issues have been both descriptively and comparatively analyzed based in the literature included in the review. This last section presents three topics: the answer to the research questions previously presented, the summary of future trends, challenges and directions, and a brief conclusion.  

\subsection{Answer to research questions}

\begin{figure}[h]
    \centering
    \includegraphics[width=0.7\linewidth]{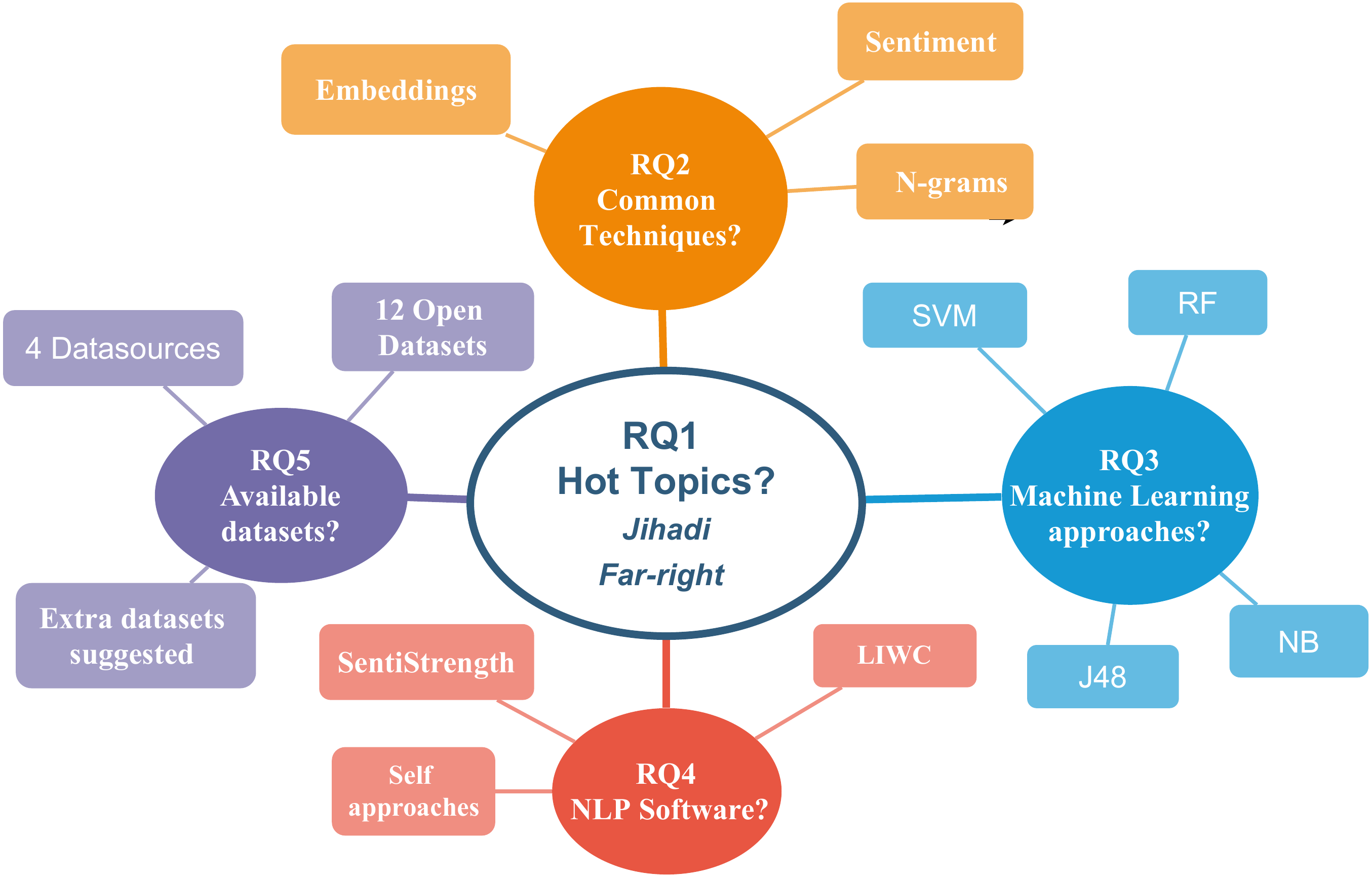}
    \caption{Research questions' answer highlights.}
    \label{fig:RQ_summary}
\end{figure}

The different research questions, regarding the state of the literature, were presented on the introduction as a justification to conduct the survey. These research questions can now be answered through the detailed analysis and insights drawn from the literature review process conducted in this article. Figure \ref{fig:RQ_summary} shows a summary of the conclusions reached after the exhaustive review, highlighting the main findings drawn for each of the questions posed at the beginning of the review. Each of these answers is explained in more detail below.

\begin{itemize}
    \item \textbf{RQ1. What are the current topics and contributions from NLP to extremism research?}
    
    Literature related to NLP approaches to extremism research has experienced a growth over the last years. Religious extremism remains as the most covered topic, followed by far-right extremism. 
    Terrorism (specially Jihadist terrorism) and counter-terrorism appear to be key motivations behind the interest for these topics, as detecting extremist content can help preventing radicalisation processes and, therefore, avoid attacks as the one experienced on recent years \cite{johansson2017detecting}. 
    
    The interest for extremism detection appears reflected on the many mentions to machine learning algorithms, as their combination with NLP approaches can be useful to create classification models to identify extremist content. Finally, even though it is beyond the scope of this review, SNA also appears as an analytical approach commonly linked to the study of language on extremism research.
    
    \item \textbf{RQ2. What NLP techniques are used on extremism research?}
    
    Section \ref{Techniques} highlights that n-grams, TF/TF-IDF and sentiment analysis are the techniques most commonly used to study extremist discourse. It is unsurprising to see the first two approaches as the most common, taking into account that they are a previous step to conduct more complex analysis, for example sentiment analysis itself. 
    
    However, it shall be considered that the use of neural networks models (word embedding) is getting more recurrent on the study of extremist discourses, and therefore it shall be considered for researchers interested on this topic. This is specially relevant, as authors have pointed that detecting a the most commons terms used in the specific domain is not enough to understand in what meaning they are being used in the text. Therefore, techniques capturing information about the context and the meaning of the terms (e.g. embedding or semantic tagging) shall also be considered as an important part of any textual analysis. Specially, taking into account that extremist texts use words from regular discourses, but with different objectives. 
 
    \item \textbf{RQ3. How have NLP techniques been applied in the field of extremism research?}
    
    54.68\% of the articles reviewed performed classification tasks using ML approaches, as stated on Section \ref{Machine}. Again, this is unsurprising, as the main objective of extremism researchers is to detect and prevent that content. Among the ML algorithms, SVM was the most commonly used, followed by Random Forests, Naïve Bayes and Decision Trees. Concerning the best models performing classification, SVM reported having a general good performance. However, in the most recent research works, Neural Networks approaches performed specially good compared to other models, and appear as a promising trend on the detection of extremism.
    
    The rest of the articles (see Section \ref{Descriptive} focused on describing the main features that differentiate between regular and extremist texts, with the interest of defining this type of discourse. This helped providing insights that could be helpful for future researchers to identify which textual features are more useful to analyse to detect (and prevent) extremism on Social Media.  
    
    \item \textbf{RQ4. What NLP software tools are commonly used on extremism research?}
    
    Section \ref{Software} highlights SentiStrength as the most used tools to conduct NLP analysis. Specifically, this tool is used to conduct sentiment scoring, through automatic tagging of words around a token. The second one is LIWC, a tool based on dictionaries with a psycholinguistic approach.
    
    Two points shall be stated here. First, 25 articles did not report the software tool they used to conduct the analysis. Second, 17 articles used a software tool used by less of three articles. Therefore, while several NLP software tools were used, it can be stated that there is not a commonly used technique in the literature to conduct NLP analysis.
    
    \item \textbf{RQ5. Which publicly available datasets or datasources have authos used to conduct NLP experiments on extremism research?}
    
    Most of the articles included on the review relied on their own private datasets to conduct their research. However, some of the datasets, specially those concerning religious radicalization and Twitter, forums or radical magazines, remain currently public. A summary of those public datasets, together with extra datasets suggested by the authors, are presented on section \ref{Datasets}. 
    
\end{itemize}

\subsection{Future trends and challenges}

\begin{figure}[h]
    \centering
    \includegraphics[width=0.85\linewidth]{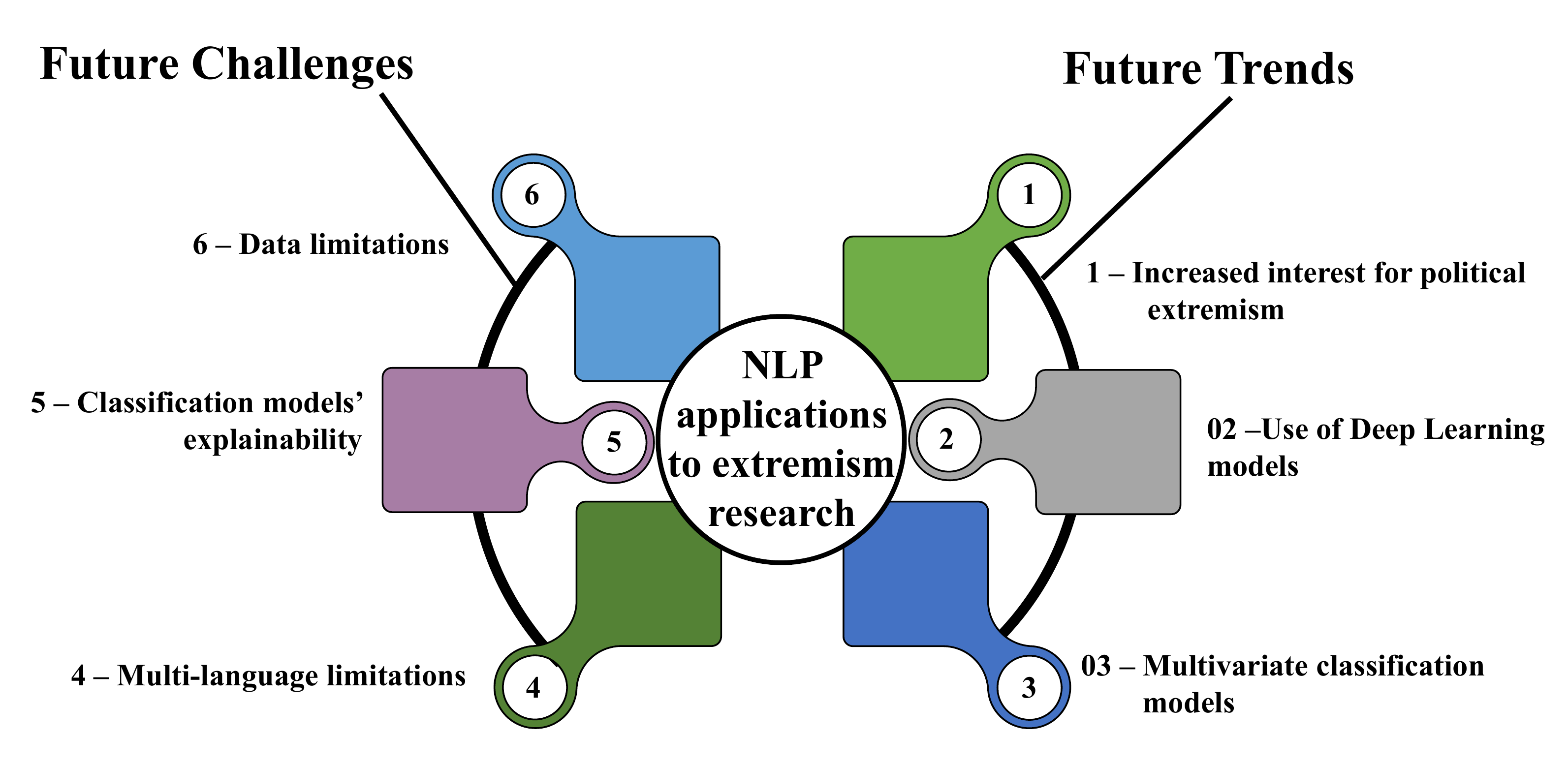}
    \caption{Future trends and challenges of NLP approaches applying on extremism research area.}
    \label{fig:future}
\end{figure}

The research questions and their answers provide a general picture of the current state of the art concerning contributions of NLP on extremism research. However, the analysis conducted on this survey also provides with different insights concerning the future of the area. This section presents the future trends that the literature will follow concerning its current state, the challenges that will be faced, and the directions to confront those challenges (see Figure \ref{fig:future}).

As shown in Figure \ref{fig:future}, there are 3 main trends that can be derived from the research questions and 3 future challenges for the NLP applications to extremism research, which are explained in detail as follows:

\begin{itemize}
    \item \textit{Future Trends}: 
     \begin{enumerate}
         \item Interest for political extremism relevance will grow in a short term. At the time this survey is being written, the Capitol assault and the shutdown of Parler (one of the most famous online bastions of Far-right groups) have attracted the interest of both the general public and researchers. In fact, several datasets concerning online political extremism are being released nowadays, which will also increase the opportunities of studying this phenomena, also taking advantage of the lessons learnt on the study of religious extremism. Therefore, this research field remains promising for future years.
    

    \item Concerning Machine Learning for extremist prediction, Neural Network based techniques have shown promising outcomes on the articles reviewed. Therefore, and due to the few literature that has approached extremist classification from this perspective, the use of these techniques remains as one promising trend for the future. The use of Deep Learning-NLP approaches (based on neural language models), also provides a way to overcome the lack of semantic information extracted from the texts, which is a key challenge in the study of extremist discourses. However, the overcome of their main limitation for this area (the explainability of the model) will represent a turning point on the use of this type of approaches, which will lead to more accurate discriminant models. 
    
    \item Multivariate classification models (those using different types of features to discriminate) achieve better results in the reviewed papers. Furthermore, the general analysis carried out in section \ref{general} shows that some of works reviewed use Social Network Analysis to pursue research studies in the area of extremism. This approach, based on the analysis of interactions among users, could be a good complement to the study of extremist dynamics on online environments \cite{camacho2020four}. Indeed, approaches combining NLP and SNA have been used in other research fields, such as fake news \cite{zhou2020survey}, and also on some articles on the extremism area \cite{torregrosa2020analyzing}, providing good results. Therefore, the application of approaches combining techniques from both areas to not only analyze extremist behaviors based on discourse (text), but also on their dynamics in online social media, will be another relevant trend to address in a short term.
    
    \end{enumerate}


 \item \textit{Future Challenges}: 
 \begin{enumerate}
 \setcounter{enumi}{3}
    \item The presence of multiple languages on an extremist text is one of the first limitation of the research area (especially, those concerning religious extremism). This limitation, very common on these type of texts, can bias an analysis depending on the techniques used (for example, dictionaries vs n-grams), which therefore implies a lot of extra interpretative or preparatory work for the researcher. The use of new approaches, such as word embedding (and, specially, those that recognize word variations) could be the right direction to follow here, together with the creation of specific lexicons for different types of extremism. 
    
    \item The explainability of the classification models is one of the most important challenges currently facing the area due to the psychological, criminological and sociological roots of extremism. The interest of detecting extremist content relays not only on the detection itself, but on the extraction of insights to understand more about the mind of extremists. With this understanding, classification can be fine-tuned, discourses can be countered, and first signs can be identified. Therefore, a balance shall be found between the accuracy of the model and its explainability. 
    
    \item The relative absence of public data sources will remain as one of the more challenging tasks to confront on extremism research. Even though there is a lot of data that can be extracted from online platforms, such as Twitter or web forums, the ethical concerns related to anonymity and the private nature of most of the data stored cause researchers to avoid sharing their datasets. This ultimately leads researchers to create new datasets each time they want to conduct a new experiment, instead of improving data already stored with new information. Therefore, creating and sharing full datasets with other researchers, always respecting the ethical steps to do it, will facilitate the access of new researchers to this field, improving the quality and quantity of the outcomes.
    \end{enumerate}
\end{itemize}

\subsection{Conclusion}
\label{Conclusion}

Currently, extremism represents a security and ideological challenge for Europe. Different kind of movements, such as jihadi terrorism and far-right groups, have changed the political and social agenda of several countries, including hot topics that are now discussed as relevant issues for those countries \cite{ali2021far}. To confront this phenomena, it is first necessary to understand the discourse, which is a reflect of the ideology of extremist groups. Only through this understanding these movements can be countered. 

NLP, with its limitations, offers technical resources to describe these discourses, together with ways of extracting insights regarding how extremists use language compared to non-extremist groups. Through the descriptive and comparative analysis of techniques, software tools, classification approaches and datasets, this survey aims to provide the reader with a global picture of the applications NLP can provide to the study of extremism. This, ultimately, will help authors to identify future research directions, relevant trends and challenges to overcome in the study of extremist discourses.

\section*{Acknowledgements}
\label{sec:acknowledgements}

This research has been supported by Ministry of Science and Education under DeepBio (TIN2017-85727-C4-3-P) and CHIST-ERA 2017 BDSI PACMEL projects (PCI2019-103623), by Comunidad Aut\'{o}noma de Madrid under S2018/ TCS-4566 (CYNAMON) and S2017/BMD-3688 grants and by the project DeepSCOP-Ayudas Fundación BBVA a Equipos de Investigación Científica en Big Data 2018. Eugenio Martínez-Cámara is supported by the Spanish Government fellowship program Juan de la Cierva Incorporación (IJC2018-036092-I). Javier Del Ser acknowledges funding support received from the Basque Government (Consolidated Research Group MATHMODE, ref. IT1294-19).

\bibliographystyle{unsrt}  
\bibliography{references}  






\end{document}